\documentclass[final,authoryear,3p,times,twocolumn]{elsarticle}
\usepackage{amssymb}
\usepackage{subfigure}
\usepackage[colorlinks=true, pdfstartview=FitV, linkcolor=black, citecolor= black, urlcolor= black, bookmarks=false, draft]{hyperref}
\usepackage{amsmath}
\usepackage[utf8x]{inputenc}	% for the polish letter in 'Skłodowska'
\usepackage{natbib}				% to show preview of \citep

\usepackage{lineno}

\begin{document}

\begin{frontmatter}

%% Title, authors and addresses

%% use the tnoteref command within \title for footnotes;
%% use the tnotetext command for the associated footnote;
%% use the fnref command within \author or \address for footnotes;
%% use the fntext command for the associated footnote;
%% use the corref command within \author for corresponding author footnotes;
%% use the cortext command for the associated footnote;
%% use the ead command for the email address,
%% and the form \ead[url] for the home page:
%%
%% \title{Title\tnoteref{label1}}
%% \tnotetext[label1]{}
%% \author{Name\corref{cor1}\fnref{label2}}
%% \ead{email address}
%% \ead[url]{home page}
%% \fntext[label2]{}
%% \cortext[cor1]{}
%% \address{Address\fnref{label3}}
%% \fntext[label3]{}

\title{The role of fragment shapes in the simulations of asteroids as gravitational aggregates}

%% use optional labels to link authors explicitly to addresses:
%% \author[label1,label2]{<author name>}
%% \address[label1]{<address>}
%% \address[label2]{<address>}

\author[PoliMi]{F. Ferrari}
\author[OCA]{P. Tanga}

\address[PoliMi]{Department of Aerospace Science and Technology, Politecnico di Milano, Italy}
\address[OCA]{Universit\`e C\^ote d'Azur, Observatoire de la C\^ote d'Azur, CNRS, Laboratoire Lagrange, France}

\begin{abstract}
	Recent remote measurements and in-situ observations confirm the idea that asteroids up to few hundreds of meters in size might be aggregates of loosely consolidated material, or `rubble piles'. The dynamics of these objects can be studied using N-body simulations of gravitational aggregation.
	We investigate the role of particle shape in N-body simulations of gravitational aggregation. We explore contact interaction mechanisms and study the effects of parameters such as surface friction, particle size distribution and number of particles in the aggregate. As a case study, we discuss the case of rubble pile reshaping under its own self-gravity, with no spin and no external force imposed.
	We implement the N-body gravitational aggregation problem with contact and collisions between particles of irregular, non-spherical shape. Contact interactions are modeled using a soft-contact method, considering the visco-elastic behavior of particles' surface. We perform numerical simulations to compare the behavior of spherical bodies with that of irregular randomly-generated angular bodies. The simulations are performed starting from an initial aggregate in a non-equilibrium state. The dynamics are propagated forward allowing particles to settle through reshaping until they reach an equilibrium state. Preliminary tests are studied to investigate the quantitative and qualitative behavior of the granular media.
	The shape of particles is found to play a relevant role in the settling process of the rubble pile aggregate, affecting both transient dynamics and global properties of the aggregate at equilibrium. In the long term, particle shape dominates over simulation parameters such as surface friction, particle size distribution and number of particles in the aggregate.
	Spherical particles are not suitable to model accurately the physics of contact interactions between particles of N-body aggregation simulations. Irregular particles are required for a more realistic and accurate representation of the contact interaction mechanisms.
\end{abstract}

\begin{keyword}
%% keywords here, in the form: keyword \sep keyword
%% MSC codes here, in the form: \MSC code \sep code
%% or \MSC[2008] code \sep code (2000 is the default)
Asteroids \sep
Asteroids, composition \sep
Asteroids, dynamics \sep
Asteroids, surfaces
\end{keyword}

\end{frontmatter}

%%\linenumbers

%% main text
\section{Introduction}
\label{s_1:intro}
	In the last few decades, remote measurements and in-situ observations have contributed to build evidence~\citep{Richardson2002,Hestroffer2019} upon the idea that asteroids up to few hundreds of meters in size might be aggregates of loosely consolidated material, or `rubble piles'~\citep{Chapman}. Due to their properties, the dynamical evolution of these objects can be studied using numerical N-body simulations. This technique has been used successfully in the past to simulate a wide range of phenomena, including planetary ring dynamics~\citep{Porco2008,Schmidt2009,Ballouz2017,Lu2018} and planetesimal dynamics~\citep{Richardson2000}. In the context of rubble-pile dynamics, N-body simulations have been crucial to provide insights on phenomena involving natural reshaping processes towards equilibrium shapes~\citep{Richardson2005, Tanga2009,Tanga2009a}, spin evolution and rotational breakup~\citep{Sanchez2012,Sanchez2016,Cotto-Figueroa2015,Ballouz2015,Zhang2017,Zhang2018,Yu2018} towards the formation of binary systems~\citep{Walsh2008}. These give insights on the internal structure of such bodies, which to date remains largely unknown~\citep{CampoBagatin2018,Sanchez2018,Scheeres2018}. The intrinsic granularity of N-body models makes them ideal candidates to investigate scenarios that involve the disruption of rubble-pile objects, and subsequent reaccumulation after catastrophic~\citep{Michel2001,Michel2002,Michel2004,Michel2013} or non-catastrophic events~\citep{Geissler1996}, to investigate the formation of satellites~\citep{Durda2004}, exposed internal structure~\citep{Benavidez2012,Benavidez2018} and interactions with external objects, including tidal disruption events~\citep{Asphaug1994,Asphaug1996,Yu2014} and low-speed collisions~\citep{Leinhardt2000,Ballouz2014}.

	All these studies consider contact and collision interactions between particles of spherical shape. The use of spheres in N-body simulations is very beneficial from the computational point of view, especially when dealing with a high number of bodies, since it reduces dramatically the complexity of the collision detection and contact solver algorithm. However, the use of spheres might affect significantly the realism of the contact interactions~\citep{Michel2004} which are over-simplified by disregarding the geometrical effects due to the irregularity and angularity of real bodies~\citep{Tanga1999,Richardson2009}. This is confirmed by experiments in terrestrial applications of granular dynamics~\citep{Pazouki2017,Dubois2018}. 
	
	Few studies have been performed in the past involving non-spherical particles. \cite{Korycansky2006} studied low-speed collisions between gravitational aggregates made of polyhedral particles and showed a substantial difference between the contact interaction mechanism between polyhedra, as compared to spheres. In particular, they highlighted how off-center collisions, which are only possible between non-spherical objects, produce significantly lower restitution at contact and consequently a higher aggregation probability after collision. In a follow-up study,~\cite{Korycansky2009} managed to obtain a substantial speed-up of their code, by employing the ODE (Open Dynamics Engine) physics engine. \cite{Movshovitz2012} use similar aggregates to investigate tidal disruption events, showing how the use of non-spherical particles produces a better estimate of the bulk density of the aggregate object compared to using spheres. Other studies make use of correction parameters applied to spherical objects to model rolling friction~\citep{Schwartz2012,Yu2014} of spheres. Compared to perfect spheres, these provide a more realistic representation of the rolling motion, and may be used to model dissipation phenomena occurring between particles. However, using a coefficient of rolling friction to model non-spherical shape is not realistic and cover only a limited range of phenomena due to the angularity of particle shape. In particular, rolling friction is dissipative only and always acts against rolling motion, whereas shape could also enhance rolling~\citep{Wensrich2012}. For the same reason, it cannot be used to reproduce accurately interlocking between particles. Also, it does not provide any means to model the dissipation of energy due to off-center collisions~\citep{Korycansky2006}, which are always towards-center when using spheres. As discussed, this would result in a substantially lower coefficient of restitution for the case of angular bodies.

	In this paper, we investigate the role of particle shape in the numerical N-body problem with gravity and contact/collision dynamics. We model particles as rigid bodies with six degrees of freedom, including translational and rotational motion. We compare and discuss the behavior of spheres against particles with randomly-generated irregular shapes. Additionally, we study the effects of physical parameters of granular media, such as particle size distribution (mono- vs poly-disperse)~\citep{Tanga1999,Tanga2009,Barnouin2019a} and resolution in terms of number of bodies involved in the simulation. The goal is to provide insights, identify general rules and key parameters that play a relevant role in the setup of a numerical simulation of gravitational aggregation. 
	
	We first introduce the case study and useful definitions in Section~\ref{s_2:problem}. The setup of numerical simulations performed is described in Section~\ref{s_3:simulations}, including a brief introduction on the code used (Section~\ref{ss_3.1:code}) and preliminary tests run (Section~\ref{ss_3.2:tests}). Section~\ref{ss_3.3:parameters} introduces the parameter space investigated, while Sections~\ref{ss_3.4:initialaggr} and~\ref{ss_3.5:finalaggr} discuss the numerical tools and routines used, respectively, to setup the simulations and retrieve the final results. Section~\ref{s_4:results} discusses the outcome of simulations and presents aggregated results. Conclusions are finally presented in Section~\ref{s_5:conclusion}.

\section{Statement of the problem}
\label{s_2:problem}
	The primary objective is to identify the role of particle shape in numerical simulations of gravitational aggregation. To this goal, we perform several numerical simulations, to explore a wide range of parameter values and dynamical scenarios. After running some preliminary tests, we perform a simulation campaign to reproduce the self-reshaping dynamics of a rubble pile object within a range of parameter values, as detailed in Section~\ref{s_3:simulations}. At the beginning of each simulation, the particles are arranged to form an elongated ellipsoidal aggregate, which is initially in a non-equilibrium state. No external force or angular momentum are applied to the particles, which are subject to their mutual gravity only. The dynamics are propagated forward until the aggregate reach a stable equilibrium condition. More detail on the initial rubble pile model and its creation can be found in Section~\ref{s_3:simulations}.

	\subsection{Definitions}
		For the sake of clarity and to have a unique reference, we provide here definitions of quantities and symbols in use throughout the paper.
		\begin{itemize}
			\item Inertial elongation $\lambda$: it is used as a measure of the elongation of the full aggregate. It is defined as the ratio between minimum and maximum moments of inertia of the aggregate ($\lambda \le 1$). The inertia tensor of the aggregate is computed taking into account for the actual inertia tensors of all particles, whether they be spheres or angular bodies.
			\item Porosity $\phi$: it is used as a measure of the void fraction inside the aggregate. It can be written in terms of the material density $\rho$ and bulk density $\rho_b$ of the aggregate as $\phi=1-\rho_b/\rho$. Its value depends on the definition of the surface to envelope the aggregate. We compute ranges of porosity, based on a minimum and a maximum volume surface, as discussed in Section~\ref{ss_3.5:finalaggr}.
			\item Packing index $\psi$: it is defined for each particle as the mean distance with its closest twelve neighbors (twelve is the number of bodies in contact in a dense spherical packing). Its value is nondimensional and normalized to the characteristic radius of the body. Distances are evaluated from center of mass to center of mass.
			\item Contact force $F_c$: it is the resulting contact force acting on a single particle.
		\end{itemize}

\section{Numerical simulations}
\label{s_3:simulations}
	This section highlights the main features of the N-body code used and introduces the rationale of the study, providing its assumptions and limiting boundaries.

	\subsection{N-body code}
	\label{ss_3.1:code}
		The numerical model is based on the GRAINS N-body code~\citep{Ferrari2017,Ferrari2020}, a rework of the multi-physics engine \textit{Project Chrono}~\citep{Tasora2016}. GRAINS handles the dynamics of particles with six degrees of freedom each (translation and rotation), as they interact through mutual gravity and contacts/collisions. Gravitational interactions are computed by means of either direct N$^2$ integration, or using a GPU-parallel implementation of the Barnes-Hut octree method~\citep{Barnes1986,Burtscher2011,Ferrari2020}. Contact and collision interactions are computed based on the actual shape model of the body, whether it be a sphere or an angular body. The shape of angular bodies is created numerically as the convex envelope of a cloud of randomly-generated points. All clouds of points share the same statistical properties of size and shape of their bounding domain: bodies have different shape but similar characteristic size and axis ratios. For an angular body, these are defined as the size and axis ratios of its bounding domain: in this case, the maximum size of the angular body is always smaller then (or equal to) its characteristic size. 
		
		Collision detection is performed into two hierarchical steps: a \textit{broad} phase, where close pairs are identified, and a \textit{narrow} phase, where contact points between irregular shapes are found precisely using a GJK algorithm~\citep{Tasora2010,Ferrari2017}. Collision detection takes advantage of a thread-parallel implementation based on the subdivision of the domain to speed up computations. 
		
		The contact dynamics are resolved by means of a force-based soft-body DEM method~\citep{Fleischmann2015}. This approach is coherent with the features of the dynamical problem we study. In the past, studies of gravitational aggregation have been performed using hard-sphere methods~\citep[and others]{Michel2001,Tanga2009a,Richardson2011} or non-smooth contact dynamics~\citep{Ferrari2017}. Both methods consider impulsive collisions between rigid bodies~\citep{Alder1959,Jean1987,Dubois2018} and are suitable to reproduce non-smooth dynamics. However, they are not accurate to reproduce smooth dynamical processes~\citep{Gilardi2002}. For these problems, soft-contact methods~\citep{Cundall1979} are best suited~\citep{Sanchez2011,Schwartz2012,Tancredi2012}.
		
		Our problem studies the motion of particles within an aggregate, as it slowly settles under its own gravitational force, before reaching equilibrium. Due to the long-lasting contact interactions between particles, addressing this problem would require a soft-contact method to consider non-impulsive smooth collisions between bodies. Accordingly, we use a force-based DEM method, which takes into account for the visco-elastic behavior of the surface material at contact. 
		 
		Normal and tangential visco-elastic actions are exchanged between bodies at each contact point: the dynamics are modeled based on a two-way normal-tangent spring-dashpot Hertzian system. Friction is also modeled at each contact point, based on a Coulomb model.
		 
		Overall, the contact interaction is set by selecting the normal/tangential stiffness and damping coefficients of the spring-dashpot systems ($K_n$,$K_t$,$G_n$,$G_t$), and the coefficients of static/dynamic friction ($\eta_s$,$\eta_d$). A more detailed description of the contact/collision methods used can be found in~\cite{Fleischmann2015} and in the documentation of Project Chrono~\citep{ChronoURL}. The code has been extensively validated in the past, and proved its capability to address accurately problems of gravitational~\citep{Ferrari2017,Ferrari2020} and granular dynamics~\citep{Mazhar2013,Mazhar2015}, including successful comparative studies with laboratory experiments~\citep{Pazouki2017}. GRAINS provides double precision accuracy and ensure satisfactorily the conservation of energy and angular momentum through collisions~\citep{Ferrari2020}.
		 
		In the context of N-body simulations of non-spherical objects interacting through mutual gravity and contacts/collisions, attempts have been made in the past using customized integration schemes~\citep{Korycansky2006} or physics/video game engines~\citep{Korycansky2009,Movshovitz2012} to simulate hard-contact interactions between polyhedral bodies. However, in these papers approximations were made to the dynamics, so that angular momentum was not sufficiently well conserved to adequately reproduce gravitational dynamics. In this context, our work is the first to use a N-body code to study the dynamics of non-spherical bodies interacting through mutual gravity and contacts/collisions, providing a sufficient level of accuracy to reproduce accurately problems of gravitational dynamics.

	\subsection{Preliminary tests}
	\label{ss_3.2:tests}
	
		\begin{figure*}[t]
			\centering
			\subfigure[\label{fig:angleofrepose_1}]
			{\includegraphics[width=0.49\textwidth]{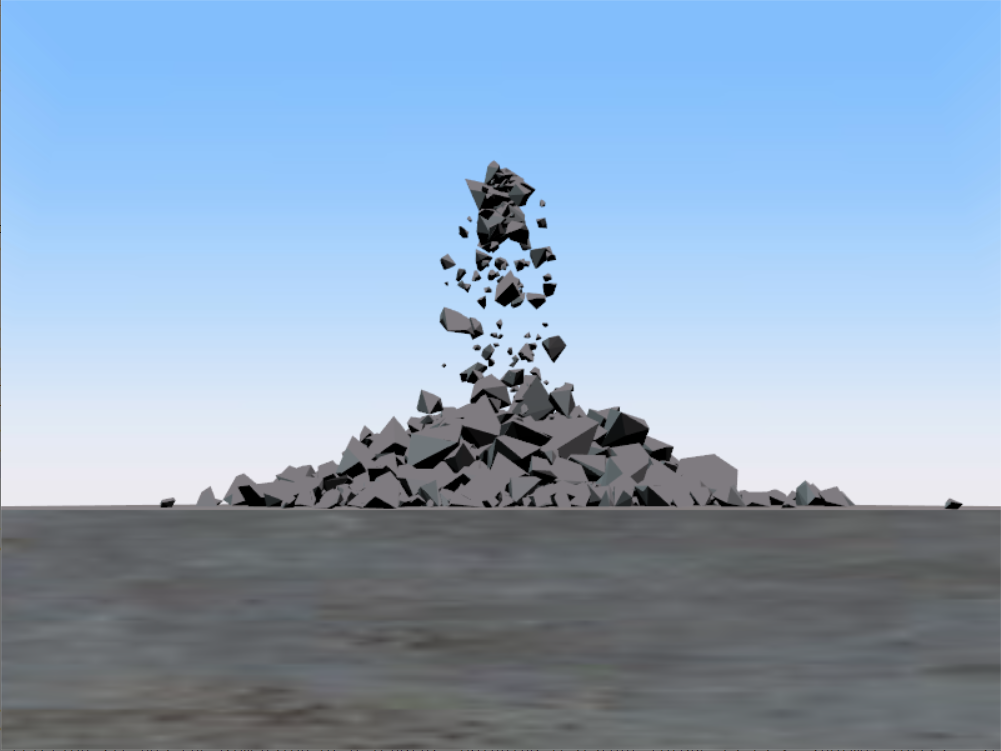}}
			\subfigure[\label{fig:angleofrepose_2}]
			{\includegraphics[width=0.49\textwidth]{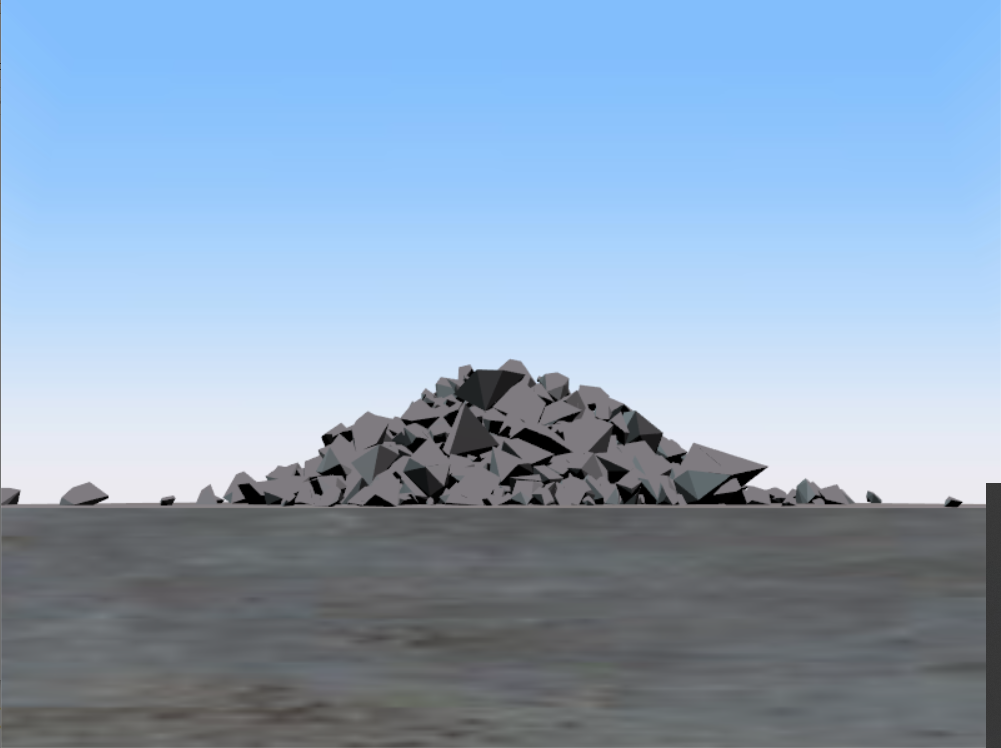}}
			\subfigure[\label{fig:angleofrepose_3}]
			{\includegraphics[width=0.49\textwidth]{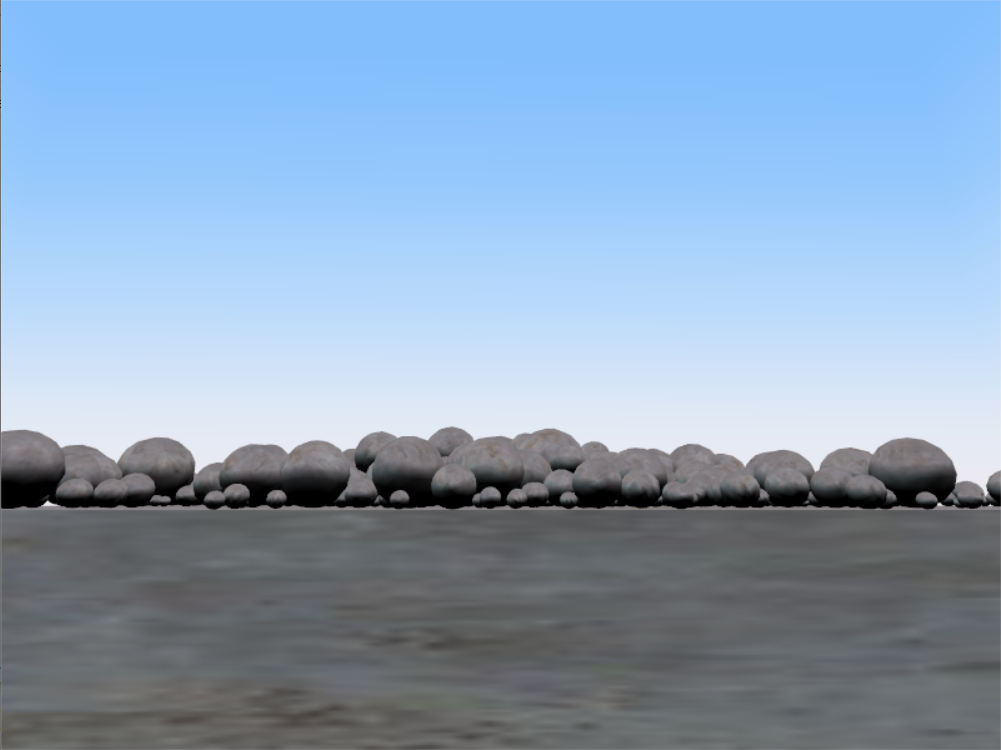}}
			\subfigure[\label{fig:angleofrepose_comp}]
			{\includegraphics[width=0.49\textwidth]{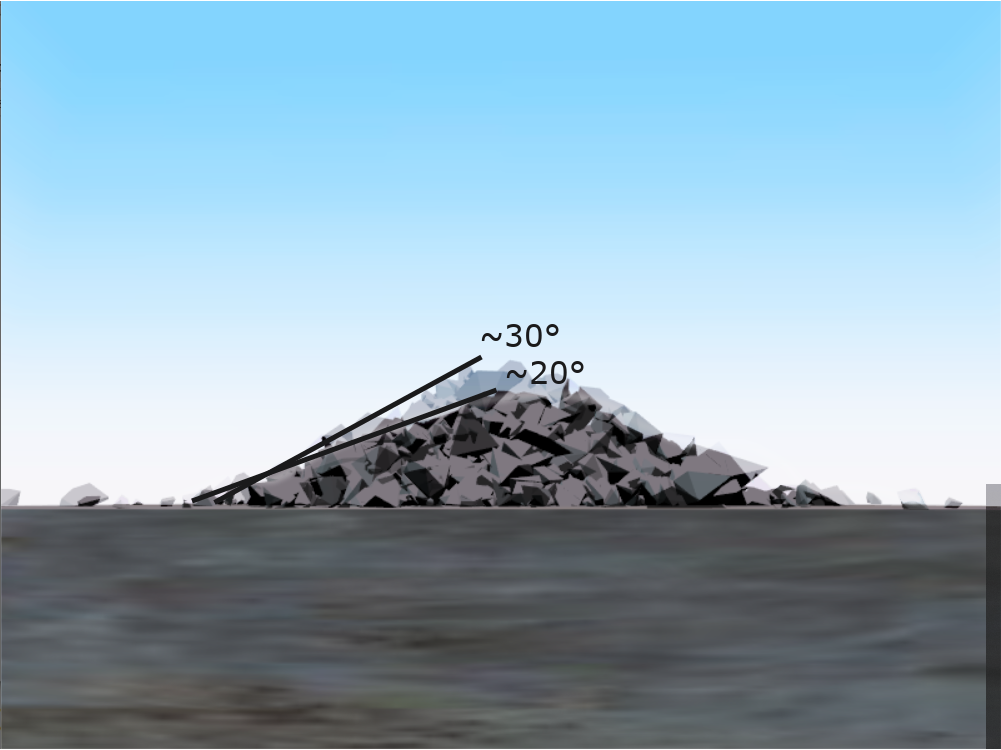}}
			\caption{Angle of repose test: (a) angular particles in the process of piling up on a flat surface; (b) pile of angular fragments at rest (coefficient of friction $\eta=0.6$); (c) spheres after the test (not piled up, $\eta=0.6$); (d) piles of angular fragments at rest: comparison between case with $\eta=0$ (slope $\simeq$20 deg, in the foreground) and $\eta=0.6$ (slope $\simeq$30 deg, in the background).}
			\label{fig:angleofrepose}
		\end{figure*}
		
		\begin{figure}[t]
			\centering
			\subfigure[\label{fig:slopes_0}]
			{\includegraphics[width=0.49\textwidth]{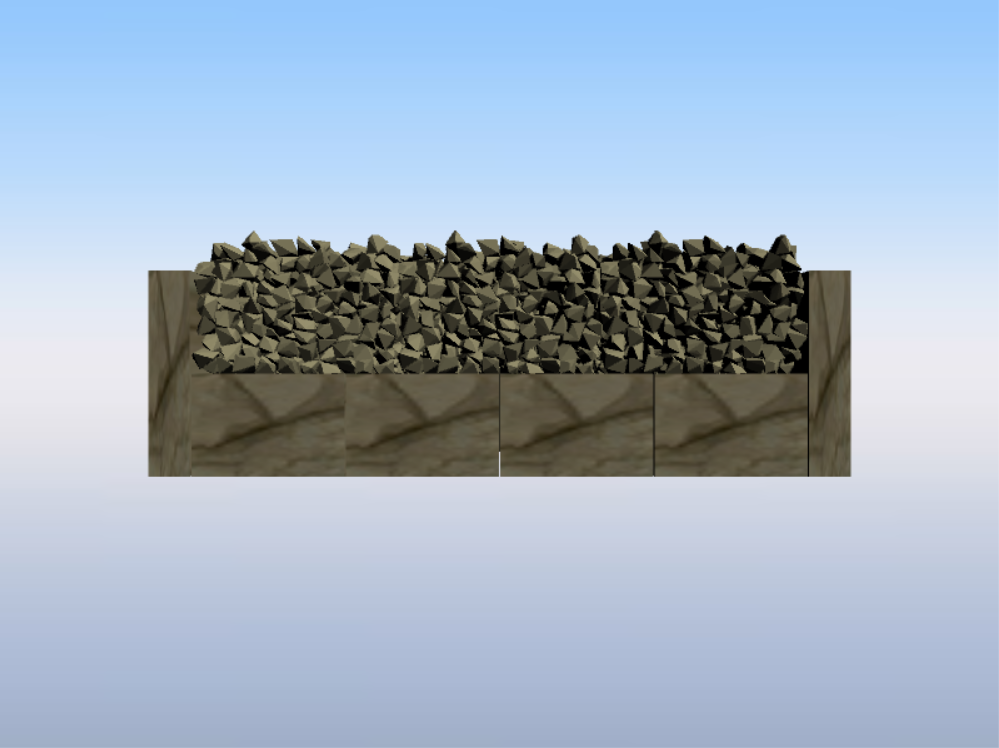}}
			\subfigure[\label{fig:slopes_1}]
			{\includegraphics[width=0.49\textwidth]{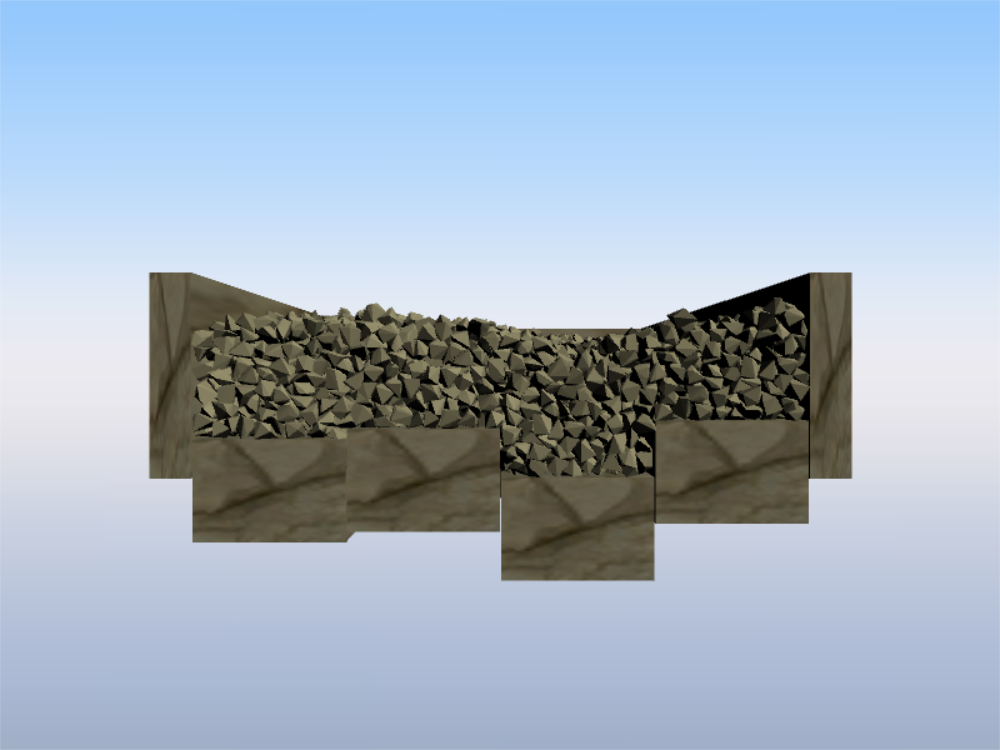}}
			\caption{Angle of slide test: creating the 3D surface. (a) The granular terrain is initially created and settled on a planar block-made floor; (b) the morphology of the granular terrain modifies after the slow downwards motion of floor blocks up to their final configuration.}
			\label{fig:slopes_creation}
		\end{figure}
	
		Before studying the full gravity/contact problem, we focus here on granular dynamics only, without considering mutual gravity between particles. As mentioned, the code is fully validated against typical granular dynamics benchmarks~\citep{Mazhar2013,Mazhar2015} and laboratory experiments~\citep{Pazouki2017}. We perform preliminary tests to complement such validation studies, with the goal to investigating the behavior of the granular media in term of angle of repose and angle of slide. In particular, we simulate the dynamics of granular terrain, as it settles dynamically (angle of repose test) and quasi-statically (angle of slide test) under a uniform gravity field. These will contribute to the discussion of the results of numerical simulations reported in Section~\ref{s_4:results}.

		\begin{figure*}[t]
			\centering
			\subfigure[\label{fig:terrain}]
%			{\includegraphics[width=0.36\textwidth]{img/terrain_noaxes}}
			{\includegraphics[width=0.49\textwidth]{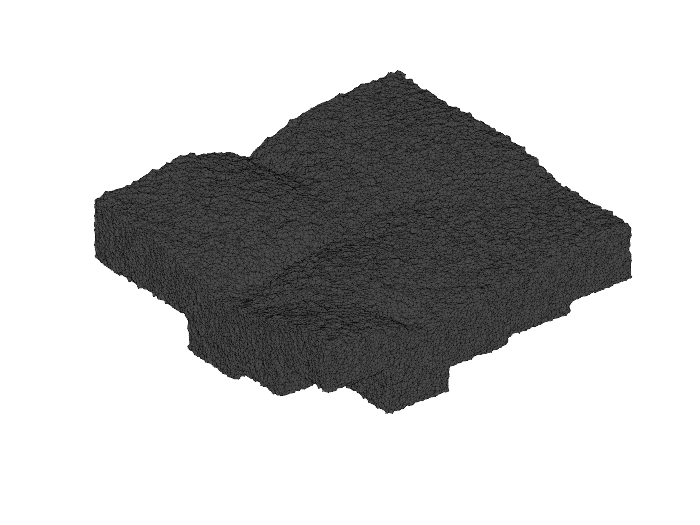}}
			\subfigure[\label{fig:terrain_points}]
%			{\includegraphics[width=0.31\textwidth]{img/terrain_points_noaxes}}
			{\includegraphics[width=0.49\textwidth]{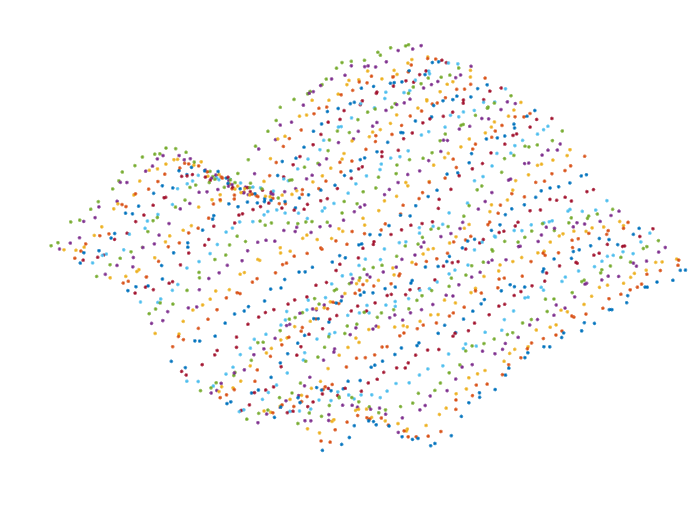}}
			\subfigure[\label{fig:terrain_mesh}]
%			{\includegraphics[width=0.31\textwidth]{img/terrain_mesh_noaxes}}
			{\includegraphics[width=0.49\textwidth]{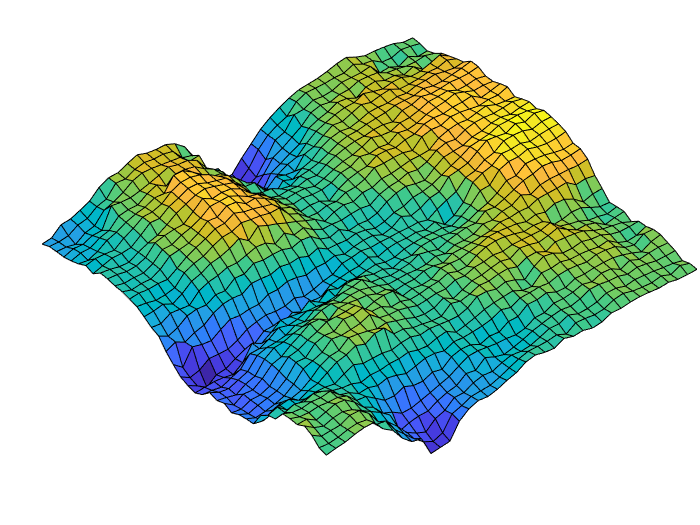}}
			\caption{Angle of slide test: finding slope distribution of the surface. (a) Enveloping surface that encloses all terrain particles; (b) grid of surface points; (c) slopes computed between points in the grid.}
			\label{fig:slopes_post}
		\end{figure*}		
		
		\begin{figure}
			\centering
			\includegraphics[width=\columnwidth]{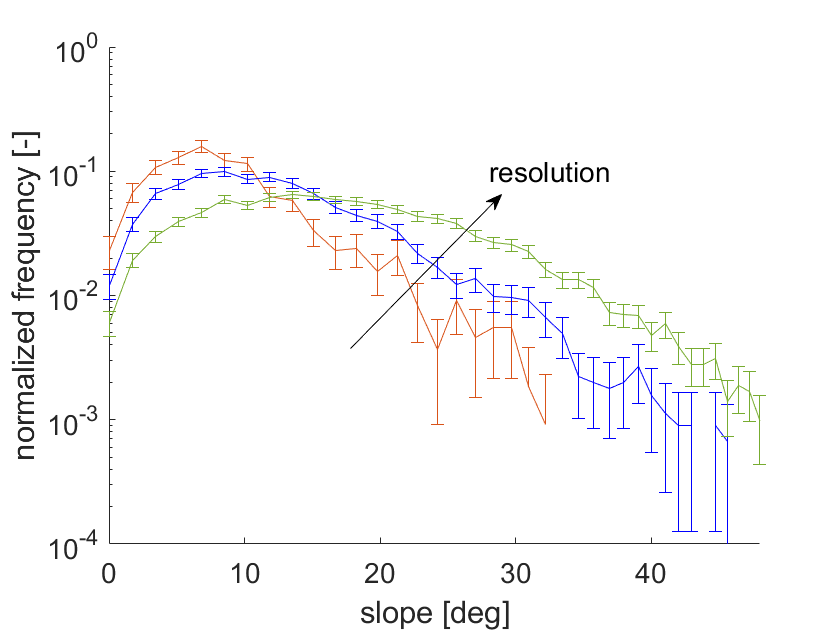}
			\caption{Angle of slide test: slope distribution examples. The higher the resolution of the model (more points in the histograms), the higher slopes can be attained.}
			\label{fig:terrain_slopes}
		\end{figure}
		
		First, we simulate the dynamics of granular material as it piles up under uniform vertical gravity, as shown in Figure~\ref{fig:angleofrepose_1}. This test gives information on the angle of repose of the granular media, i.e.\ the slope at which a landslide stops and the pile comes to rest. In our test, 3,000 particles fall from a height of 4 m on a planar surface. We use a polydisperse size distribution of particles, with average characteristic size (or diameter for the case of spheres) of 30 cm (see Section~\ref{ss_3.3:parameters} for details on the distribution function used). The coefficient of friction is set to $\eta=0.6$. 
		
		Figure~\ref{fig:angleofrepose_2} shows that angular particles reach a non-zero angle of repose, at a value of about 30 deg. This result is consistent with the coefficient of friction selected, since $\arctan(\eta)\simeq 30$ deg. On the other hand, Figure~\ref{fig:angleofrepose_3} shows that spheres are not able to pile up dynamically, showing a zero angle of repose. As shown in \cite{Walsh2008}, spheres exhibit a non-zero angle of repose when initially arranged in a compact crystallized configuration, but are not able to attain such configuration dynamically. Finally, Figure~\ref{fig:angleofrepose_comp} shows the comparison between angle of repose attained by angular particles with no friction ($\eta=0$) and the same case with $\eta=0.6$. This shows that, even with no surface friction, angular particles are able to attain a non-zero angle of repose, thanks to purely geometrical effects related to their shape (e.g.\ interlocking). In this simple example, the effects of particle shape are quantified to an angle of repose of about 20 deg, not far from the 30 deg obtained with $\eta=0.6$.

		The second test we performed is used to estimate the maximum angle of slide attainable by our granular media with angular particles. Unlike the angle of repose, which is attained dynamically, the angle of slide is attained statically, as the maximum slope that a granular material can withstand before landsliding occurs. To estimate this quantity, we simulate the slow dynamical modification of the morphology of a granular terrain. As shown in Figure~\ref{fig:slopes_creation}, we simulate a granular terrain made of angular bodies, which lies initially on a planar floor. The terrain is settled under uniform vertical gravity until reaching equilibrium. The floor is made of several cubic blocks, which can be moved independently and whose interstitial distance is much lower than the characteristic size $l_p$ of the terrain particle. The side length $l_c$ of the cubic blocks is chosen such that $l_c>10 \ l_p$. After settling, each block (subscript $j$) moves downwards with a velocity $v_j=h_j/t_{\text{sim}}$, where $t_{\text{sim}}$ is the simulation time (equal for all blocks) and $h_j$ is the final vertical displacement of block $j$. The value for $h_j$ is chosen randomly, with a uniform distribution between 0 and $l_c$, which is the maximum vertical displacement allowed. The downwards velocity of the blocks is very slow, to guarantee a quasi-static settling of the terrain during the whole simulation. In particular, the parameters of the simulation are chosen such that the value of $v_j$ is much lower than the free fall velocity of particles. In our simulations, we use 25,000 to 100,000 equally sized particles, and floors made of either 4-by-4 or 5-by-5 blocks. Inter-particle coefficient of friction is set to $\eta=0.6$ for all simulations. Maximum computation times (for the case of 100,000 particle simulation, with a 24 thread-parallel architecture), are in the order of one week.
		
		Once settled to its final configuration (Figure~\ref{fig:terrain} shows the envelope of all particles at the end of the simulation), the morphology of the terrain is acquired by means of the following procedure. The upper surface is sampled, in order to acquire surface points within a user-defined grid (Figure~\ref{fig:terrain_points}). The sampling grid is consistent with the size of the particle and typically each grid patch contains about 4 to 16 particles (2-by-2 up to 4-by-4). The highest particle point within each patch is the sampling point. The three-dimensional slopes on the surface are computed as the gradient between sampling points (Figure~\ref{fig:terrain_mesh}). The procedure of sampling patches instead of taking directly vertices or positions of fragments, is implemented to filter out slopes related to the shape of the single particle, and to acquire the granular morphology of the terrain only. 
		
		Figure~\ref{fig:terrain_slopes} shows examples of slope distributions obtained. The graph shows that higher slopes are attained in higher resolution models (more sampling points/particles in the simulation). For a lower number of particles and higher slopes, the measure points in the distribution histogram become too few and the uncertainty increases. The curves are mostly accurate in the range 0-30 degrees, where they match nicely the qualitative and quantitative behavior of measured slope distributions of small bodies~\citep[e.g.\ Fig.1 of][for the case of Vesta and Ceres]{Ermakov2019}. 
		
		In a future investigation, we plan to increase the number of bodies in the simulation, to investigate the slope distribution curve beyond 30 deg, which is currently out of the scope of this work. Although not providing accurate information on the higher slope distribution, our simulations do provide meaningful information on the maximum slopes attainable by the granular medium, which are observed in a wide range between 35 and 65 deg. As expected, these are higher than the angle of repose~\cite[e.g.][report a typical increment of about 5-10 deg over the angle of repose]{Harris2009}, which is obtained dynamically.

	\subsection{Parameter space investigated}
	\label{ss_3.3:parameters}
		The goal is to study the dynamical behavior of a rubble pile aggregate under its own gravity, as it evolves from a non-equilibrium initial state to a self-achieved equilibrium condition. To investigate the role of particle shape in the dynamical process of adjustment towards equilibrium, we performed an extensive simulation campaign, including several sets of simulations. In order to compare results of such simulation sets and to have comparable dynamics in terms of characteristic times, we use initial aggregates with the same bulk properties. For all cases, we use the same total mass, bulk density and overall shape. In particular, the initial aggregate is shaped as a prolate ellipsoid (a$>$b=c), with b/a=c/a=0.4. More details on the initial aggregates used are provided in Section~\ref{ss_3.4:initialaggr}.

		To perform comparative analyses on simulation parameters, we investigate the effects of shape, size distribution, number of particles and coefficient of static/dynamic friction. More details are provided here.
		\begin{itemize}
			\item Shape: spheres vs angular particles. Angular particles are created as the convex envelope of 15-20 points created randomly using a uniform distribution within a cubic box. Their aspect ratio is therefore close to one. Such angular bodies have on average 10 vertices.
			\item Particle size distribution: mono vs polydisperse. When mentioned, monodisperse distribution refer to particles with same radius (spheres) or characteristic size (angular particles). In all polydisperse cases, the particle size distribution is set using the Zhang probability distribution function~\citep{Zhang1999}:
			\begin{equation}
				P(s)=e^{-\frac{s-s_{\text{m}}}{\bar{s}-s_{\text{m}}}}	\quad \text{with } s\ge s_{\text{m}}
			\end{equation}
			In this case, the distribution function can be set by selecting the ratio between the average size $\bar{s}$ and the minimum size $s_{\text{m}}$. In our simulations, we use a value of $\bar{s}/s_{\text{m}}=2.70$, which corresponds to the so-called modified Zhang distribution~\citep{Elek2008}. The Zhang distribution comes with the important assumption that a minimum particle size exists in the fragment population. This is very convenient when performing numerical simulations. Also, this assumption has proven consistent with the mechanics of fragmentation~\citep{Zhang1999}, and show good agreement with laboratory experiments~\citep{Wesenberg1977,Grady1983}. These have proven the accuracy of the Zhang distribution to match size distribution after fragmentation processes~\citep{Zhang1999,Elek2008}.
			\item Number of particles: ranging between 1,000 and 2,000.
			\item Coefficient of friction: ranging from 0 to 1. In our simulations, to reduce the amount of free parameters, the dynamic coefficient of friction $\eta_d$ is always set equal to the static coefficient of friction $\eta_s$: for this reason, from now on, we use the term $\eta$ to indicate both.
		\end{itemize}
		Additional parameters include coefficients related to the surface/contact interactions. In our simulations, both normal and tangential stiffness coefficients are set to $K_n=K_t=2 \times 10^5$~N/m, while the damping coefficients are set to $G_n=20$~Ns/m and $G_t=40$~Ns/m. Values chosen are commensurate to the case study, and are within the range of values used in previous works: as reported by \cite{Sanchez2011}, the $K_t/K_n$ ratio is typically between 2/3 and 1 for most materials~\citep{Mindlin1949}. Moreover, \cite{Silbert2001} reports that  ``the contact dynamics are not very sensitive to the precise value of this ratio'' and thus the choice of these values is not critical for our simulations.

		From the numerical point of view, the dynamics of the system are propagated forward in time using a symplectic semi-implicit Euler integrator~\citep{AnitescuTasora,Mangoni2018}. The time step is selected consistently with characteristic times of the dynamics involved: gravitational dynamics, collision detection and contact dynamics. Gravitational dynamics is typically very slow and in our case its characteristic time can be estimated to be no lower than:
		\begin{equation}
			T_g\simeq\frac{1}{\sqrt{G\rho}}>10^3 \text{ s}
		\end{equation}
		where $\rho$ is the material density of particles and $G$ is the universal gravitational constant. Contact/collision dynamics is typically much faster. 
		
		The time stepping must be fast enough to avoid missing any collision and to properly reproduce the visco-elastic behavior of the material at contact. Equation~\eqref{eq:T_colldet} provides an estimate of the characteristic time between two consecutive collisions $T_d$ considering the limiting case of grazing collisions between spherical particles of radius $R$~\citep{Sanchez2011}, with relative velocity $v$ and maximum overlapping allowed $\delta$. In our simulations the settling dynamics is very slow and the relative velocity $v$ between particles is always below 1~m/s. We choose this value ($v=1$ m/s) as a limiting value to estimate the shortest characteristic time $T_d$ in the system. The maximum overlapping allowed is chosen as $\delta=R/100$, which proved to be consistent with parameters in our simulation, as demonstrated by previous similar tests~\citep{Ferrari2020}.
		
		Given the range of parameters in use, we can estimate characteristic times for collision detection and contact dynamics and derive a lower-bound constraint, respectively as~\citep{Ferrari2020}:
		\begin{align}
			\label{eq:T_colldet} T_d\simeq\frac{2\sqrt{4R\delta-\delta^2}}{v}> 10 \text{ s}	\\
			\label{eq:T_cont} T_k\simeq\frac{2\pi}{\omega_k}=2\pi \sqrt{\frac{m_r}{k}}>10^2 \text{ s} 
		\end{align}
		where $m_r$ is the reduced mass between the two least massive particles of the system (in our case the tightest constraint applies on simulations with polydisperse size distribution) and $k$ is the highest stiffness in the system (in our case is $K_n=K_t$). With the parameter values chosen for our case study, the tightest constraint on the time step is derived from collision detection characteristic time $T_d$. Accordingly, time steps are chosen in the order of few seconds.
		
		The maximum simulation time is set to 250~h (real world time) for all simulations, in order to leave enough time for all bodies to settle after reaching an equilibrium state.

	\subsection{Creation of initial aggregate}
	\label{ss_3.4:initialaggr}

		\begin{figure*}[ht]
			\centering
			\includegraphics[width=\textwidth]{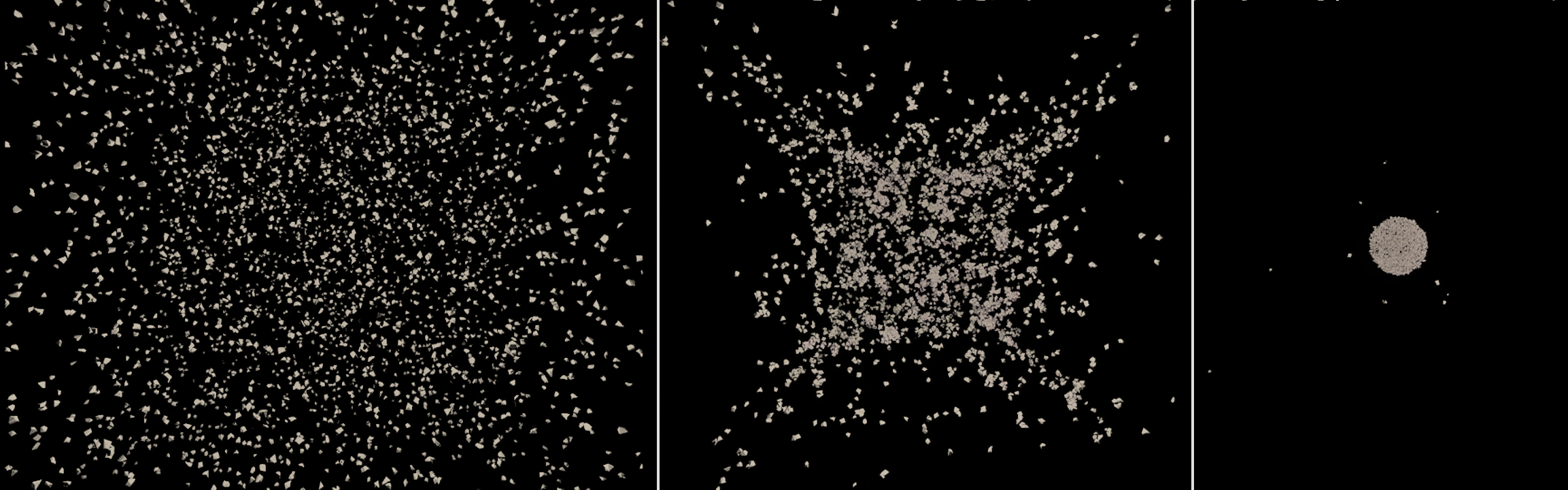}
			\caption{Snapshots from the simulation of gravitational aggregation to form a 5,000-body parent aggregate (monodisperse particle size distribution).}
			\label{fig:parent_sim}
		\end{figure*}
		
		\begin{figure}[ht]
			\centering
			\subfigure[\label{fig:CG_3dmodel}]
			{\includegraphics[width=0.49\textwidth]{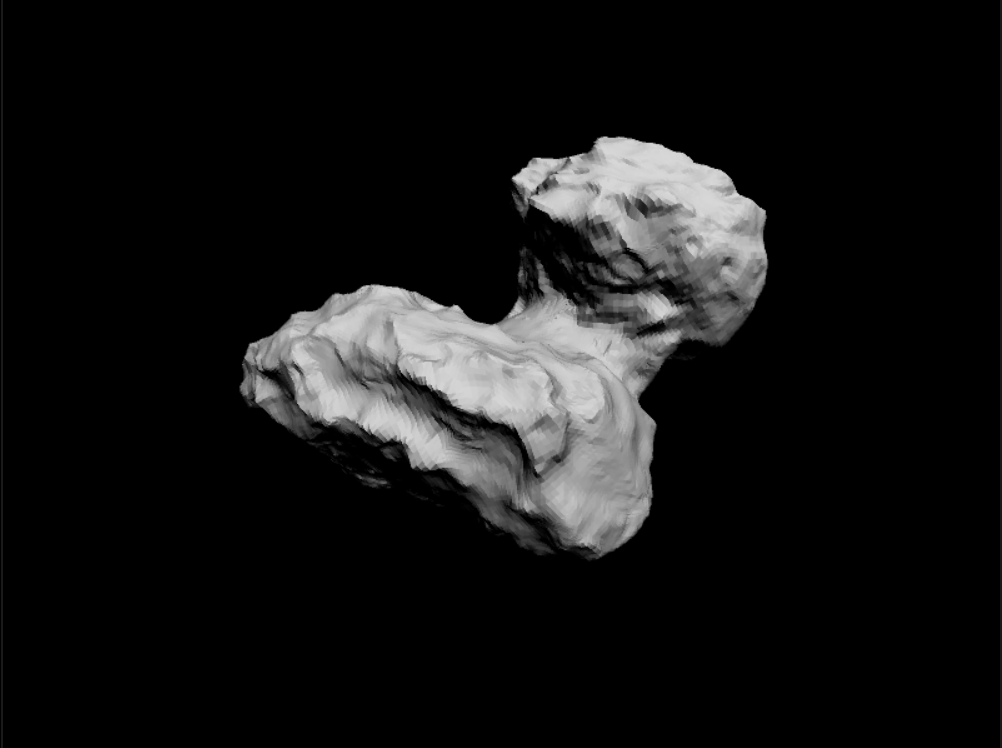}}
			\subfigure[\label{fig:CG_aggregate}]
			{\includegraphics[width=0.49\textwidth]{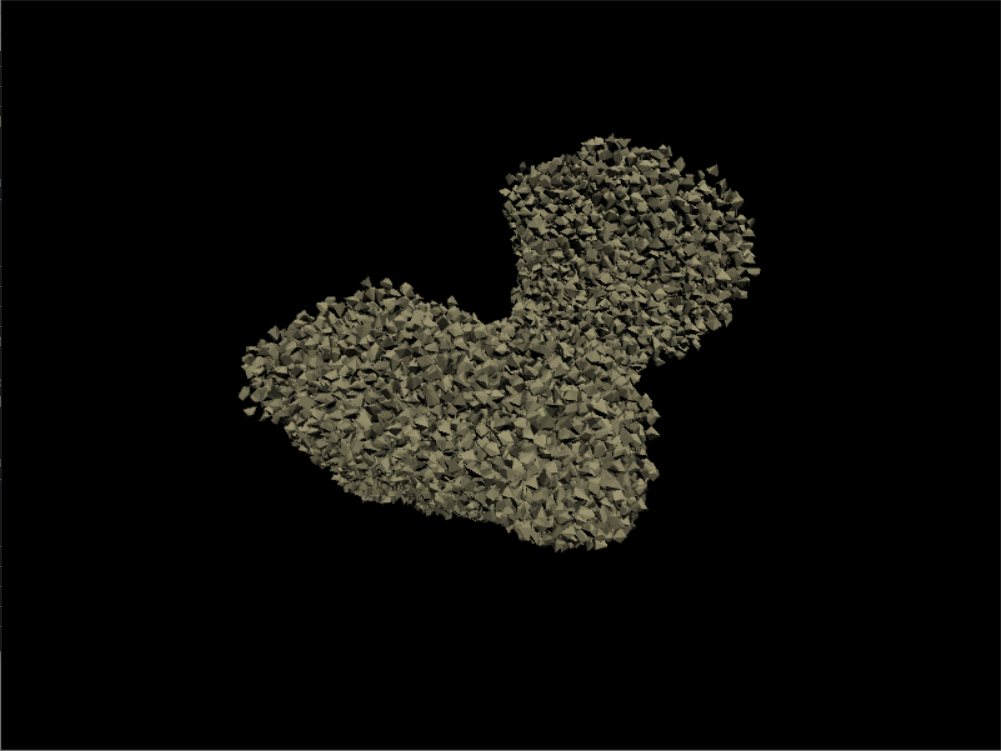}}
			\caption{Testing the shape creation algorithm using the shape of comet 67P/Churyumov-Gerasimenko: (a) 3D shape model vs (b) rubble-pile model (10,000-body model extracted from a 64,000-body parent aggregate).}
			\label{fig:CG_test}
		\end{figure}	
		
		\begin{figure}[ht]
			\centering
			\subfigure[\label{fig:bigagg}]
			{\includegraphics[height=3.6cm]{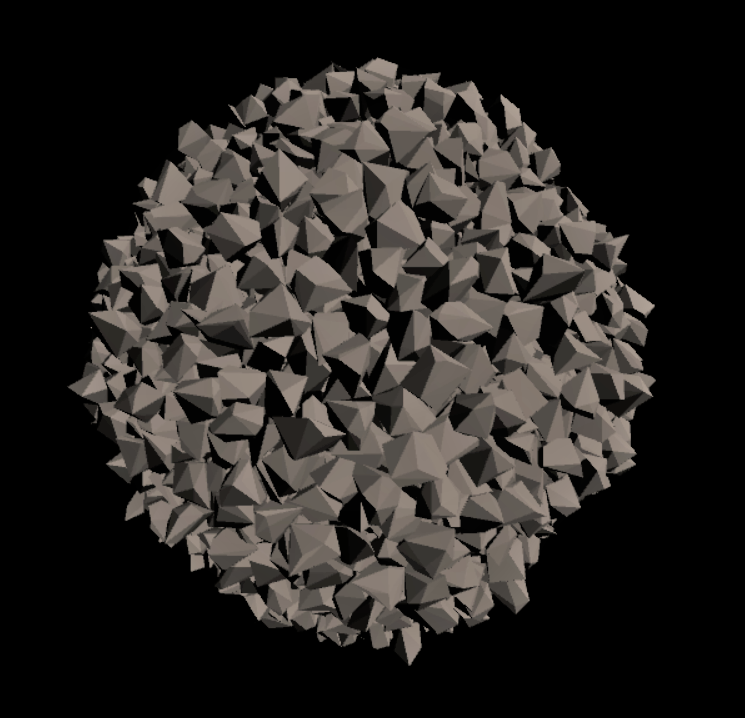}}
			\subfigure[\label{fig:extracted}]
			{\includegraphics[height=3.6cm]{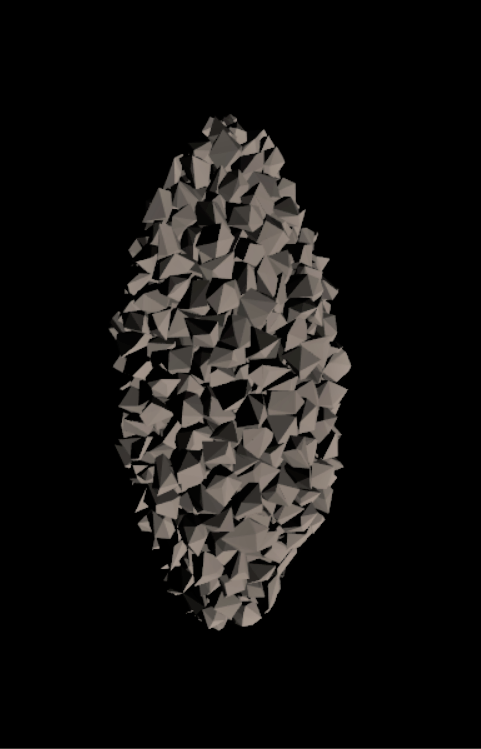}}
			\subfigure[\label{fig:bigagg_poly}]
			{\includegraphics[height=3.6cm]{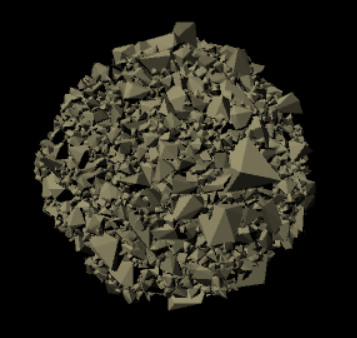}}
			\subfigure[\label{fig:extracted_poly}]
			{\includegraphics[width=3.6cm,angle=90]{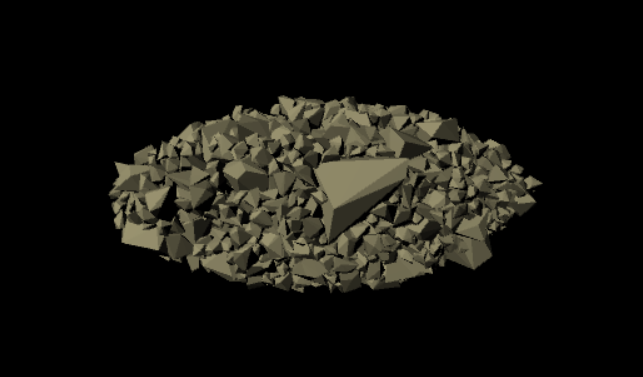}}
			\caption{Examples of initial aggregate creation. (a) 5,000-body parent vs (b) 1,000-body extracted aggregate (monodisperse particle size distribution). (c) 10,000-body parent vs (d) 2,000-body extracted aggregate (polydisperse particle size distribution).}
			\label{fig:bigextr}
		\end{figure}
		
		The initial aggregate is obtained as the result of a process of gravitational aggregation. The purpose of this is to have realistic rubble pile models with internal voids randomly distributed between irregular fragments. The gravitational aggregation process is simulated for a high number of bodies, created at random positions in a cubic physical domain, with zero velocity and spin rate, and therefore with no orbital angular momentum. After settling under self-gravity, the bodies reach a stable and nearly-spherical aggregate. Figure~\ref{fig:parent_sim} shows snapshots from the gravitational aggregation simulation (in this example to form a 5,000-body \textit{parent} aggregate). The rubble pile used in our simulations is extracted from the parent aggregate using a ray-tracing algorithm, which helps identifying particles that are inside a given 3d mesh. We validated this \textit{shape extracting} algorithm by testing it on complex and non-convex three-dimensional shapes. Figure~\ref{fig:CG_test} shows a test performed using the shape of comet 67P/Churyumov-Gerasimenko (also referenced as 67P/C-G in the followings): in this case a 10,000-body rubble pile model of the comet is extracted from a 64,000-body parent aggregate of monodisperse angular bodies. Figure~\ref{fig:bigextr} shows two ellipsoidal aggregates used in our simulations as initial rubble pile models, and their parent aggregates. In this example, a 1,000-body monodisperse rubble pile model (Figure~\ref{fig:extracted}) is extracted from a 5,000-body parent aggregate (Figure~\ref{fig:bigagg}), and a 2,000-body polydisperse model (Figure~\ref{fig:extracted_poly}) is extracted from a 10,000-body parent aggregate (Figure~\ref{fig:bigagg_poly}). The process of creating the initial aggregate is carried both for the case of spheres and angular particles. As already mentioned, the extracted aggregate is a prolate ellipsoid (a$>$b=c), with b/a=c/a=0.4, for all cases.

	\subsection{Identification of final aggregate}
	\label{ss_3.5:finalaggr}
\begin{figure}[t]
			\centering
			\subfigure[\label{fig:CG_alphashape}]
			{\includegraphics[width=0.49\textwidth]{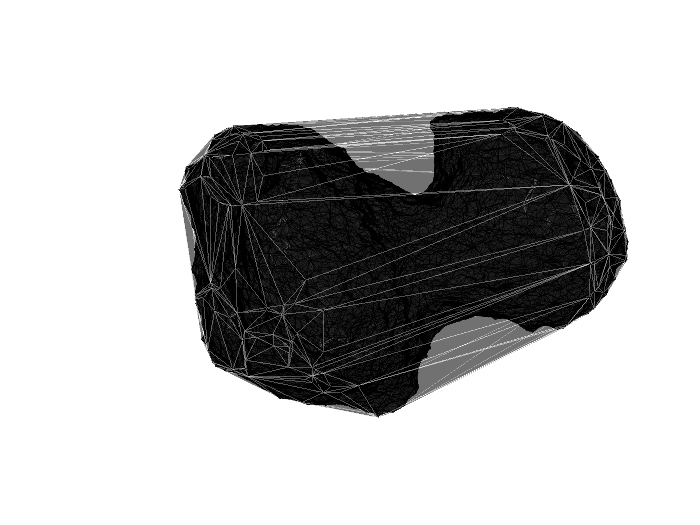}}
			\subfigure[\label{fig:alphashape_ell_min}]
			{\includegraphics[height=0.22\textwidth,angle=90]{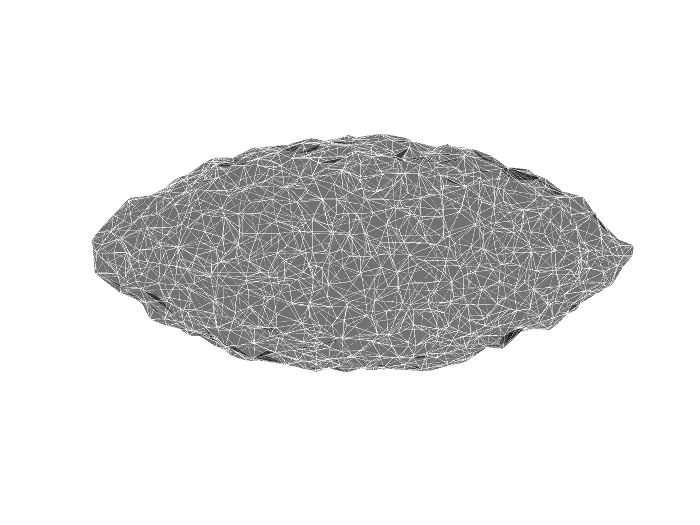}}
			\subfigure[\label{fig:alphashape_ell_conv}]
			{\includegraphics[height=0.22\textwidth,angle=90]{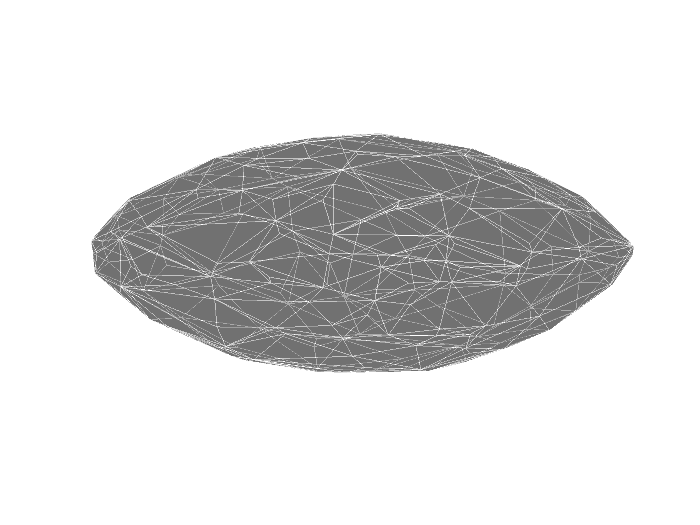}}
			\caption{(a) Testing the alpha-shape algorithm using the shape of comet 67P/Churyumov-Gerasimenko: minimum volume shape vs convex envelope. Examples of (b) minimum volume shape and (c) convex envelope for the case of ellipsoidal aggregate.}
			\label{fig:alphashape}
		\end{figure}

		The identification of the final aggregate in terms of number of bodies and mass is straightforward. Conversely, the identification of its overall shape is arbitrary and relies on the definition of its external surface. This definition is non-unique, given the information available (shape and position of each particle in the aggregate). This arbitrary process affects the computation of global properties such as volume and porosity. To provide a proper estimate of global properties, we provide ranges of values for these quantities, referring to two limiting surfaces: a maximum and a minimum volume surface. The maximum volume surface is the convex envelope of the aggregate, which is unique, when considering all vertices of angular bodies or the external surface of spheres. The minimum volume surface is computed using an alpha-shape algorithm~\cite{Edelsbrunner}: intuitively, the alpha-shape envelope is the surface created by a sphere of radius similar to the characteristic radius of particles, as it rolls over the cloud of points, whether they be the vertices of angular bodies or the external surface of spheres. To test the ability of the alpha-shape algorithm to identifying concave enveloping surfaces, we use it to find the minimum volume surface of comet 67P/C-G, starting from its rubble-pile model shown in Figure~\ref{fig:CG_aggregate}. Using this method, the surface of the comet is obtained very accurately, within the resolution given by the characteristic size of particles. Figure~\ref{fig:CG_alphashape} shows the minimum volume surface (black surface), together with the convex envelope of the comet (light gray surface). In this case, since 67P/C-G is non-convex, its convex envelope does not reproduce accurately its shape. However, in our simulations, all aggregates are ellipsoid (convex) and none of them ever reach a non-convex shape during (or at the end) of the simulation. Therefore, in our case, it makes sense to use the convex envelope to set an upper limit, as a reliable measure of the maximum volume surface the aggregate. Figures~\ref{fig:alphashape_ell_min} and~\ref{fig:alphashape_ell_conv} show examples of, respectively, the minimum and maximum volume surfaces of an ellipsoidal aggregate. In this case the differences are limited to the roughness of the envelope's surface and appear to be minimal. Still, the change in volume (and therefore in bulk density or porosity, since the total mass is constant) can be significant.

\section{Simulations of self-gravitational settling}
\label{s_4:results}
	The results of the simulation campaign are discussed here. In particular, we highlight how the shape of the particles affects the properties of the aggregate during its dynamic evolution and after the transient phase, when reaching its final equilibrium state.
	
	As described in Section~\ref{ss_3.4:initialaggr}, the aggregate used in these simulations is extracted from a larger parent, which is the result of a gravitational aggregation process. This child aggregate (also referenced as initial aggregate in the followings) is elongated and thus not in equilibrium: its self-gravitational settling is studied here.

	\subsection{Time evolution of the aggregate}

		\begin{figure*}[ht]
			\centering
			\subfigure[\label{fig:FvsCH_z}]
			{\includegraphics[width=0.49\textwidth]{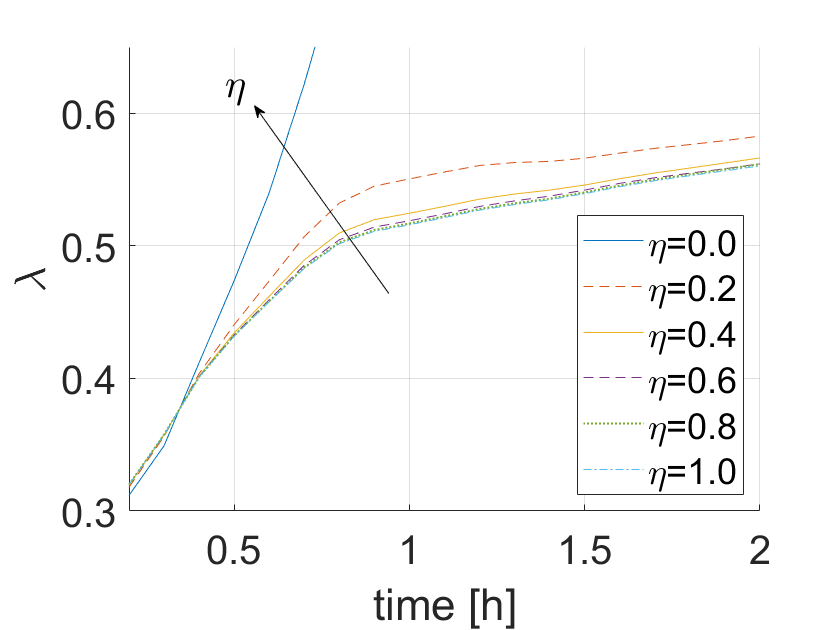}}
			\subfigure[\label{fig:FvsCH}]
			{\includegraphics[width=0.49\textwidth]{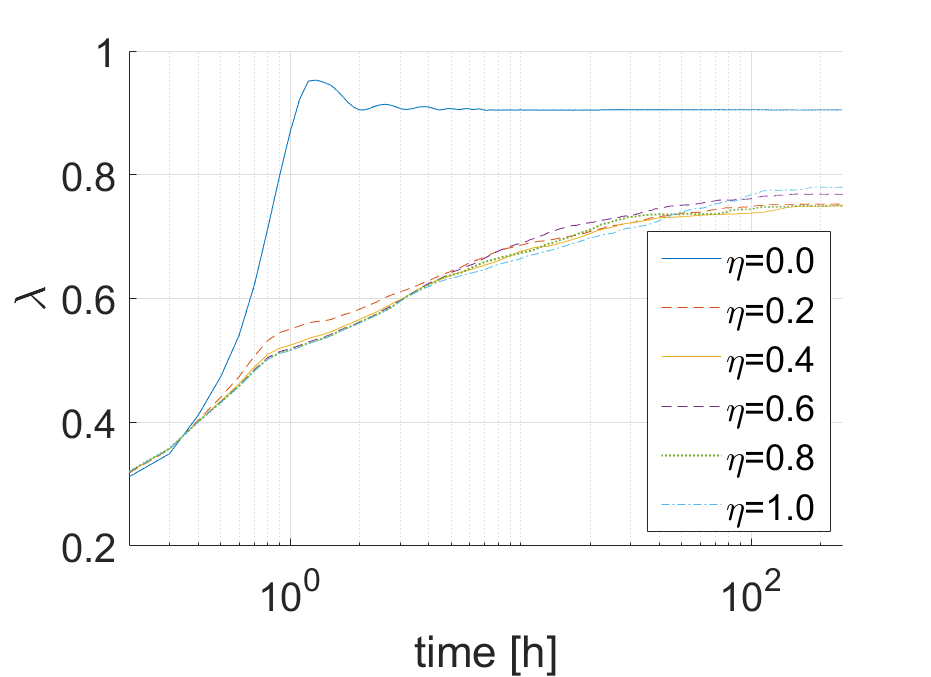}}
			\subfigure[\label{fig:FvsSP_z}]
			{\includegraphics[width=0.49\textwidth]{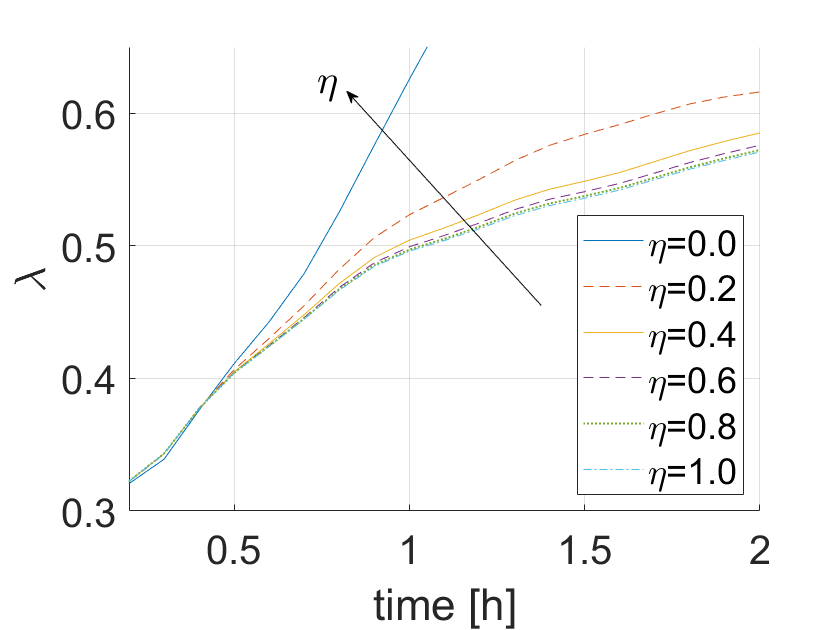}}
			\subfigure[\label{fig:FvsSP}]
			{\includegraphics[width=0.49\textwidth]{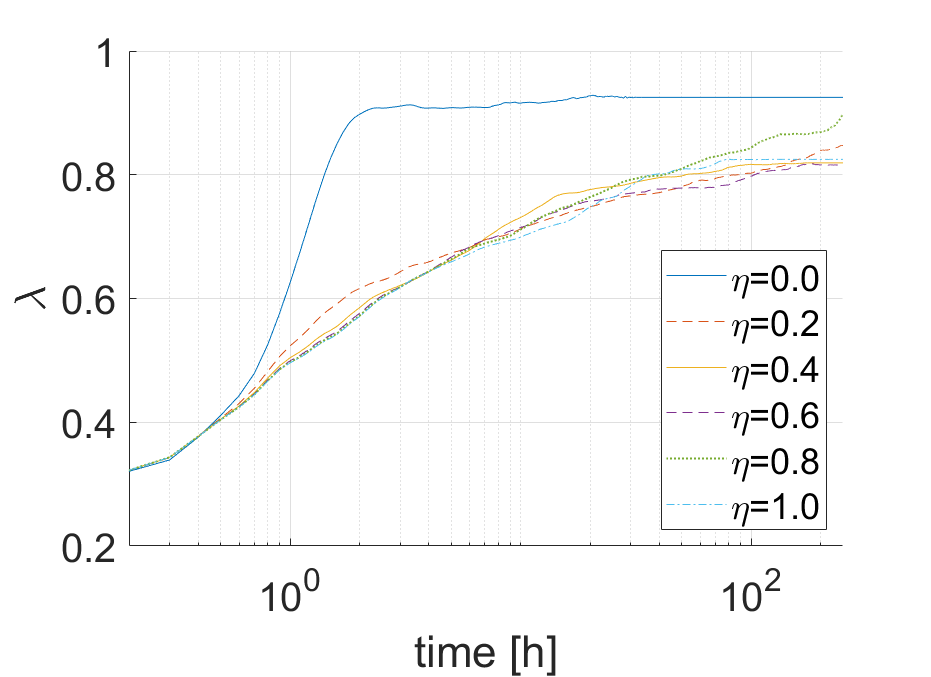}}
			\caption{Short- and long-term evolution of inertial elongation $\lambda$ of the aggregate, as function of friction coefficient $\eta$, for (a,b) angular bodies and (c,d) spheres. Long-term evolution plots (b,d) are shown using a semi-logarithmic (x) scale. The curves for $\eta=0$ are out of scale in short-term plots: after 2 hours $\lambda\simeq0.9$ in both (a) and (c) plots. Most of the simulations reach a static shape within the simulation interval.}
			\label{fig:FvsShape}
		\end{figure*}
		
		\begin{figure*}[ht]
			\centering
			\subfigure[\label{fig:iSR_mono_z}]
			{\includegraphics[width=0.49\textwidth]{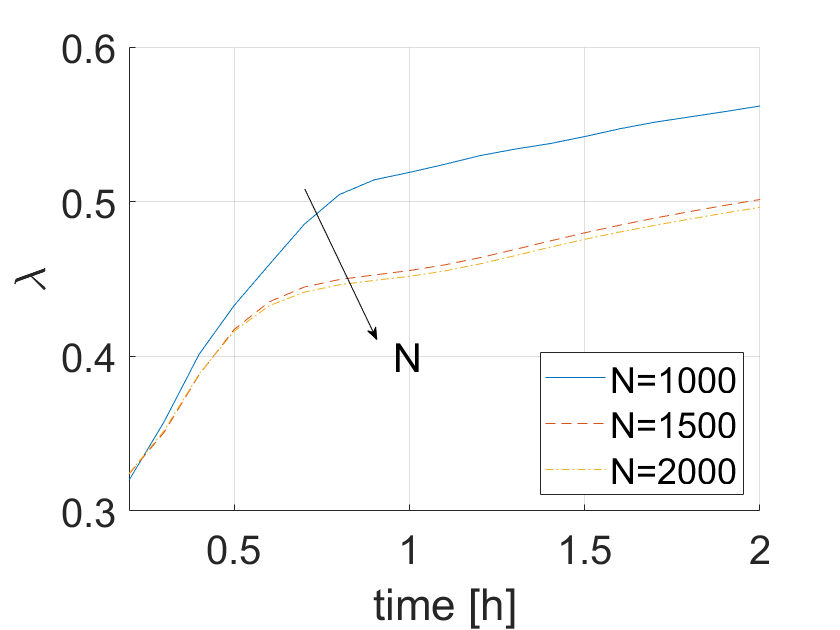}}
			\subfigure[\label{fig:iSR_mono}]
			{\includegraphics[width=0.49\textwidth]{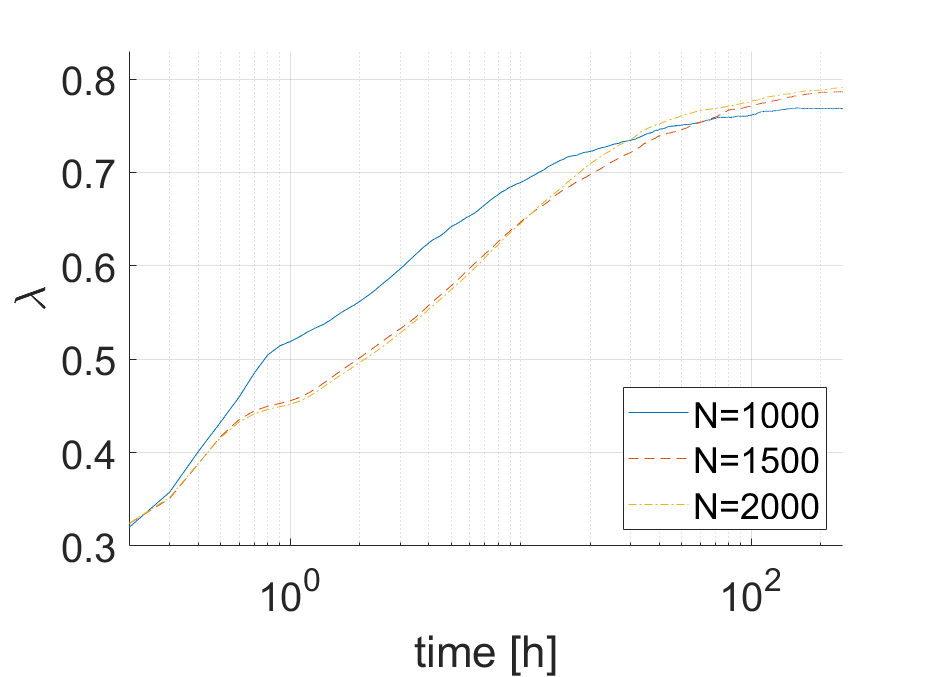}}
			\subfigure[\label{fig:iSR_poly_z}]
			{\includegraphics[width=0.49\textwidth]{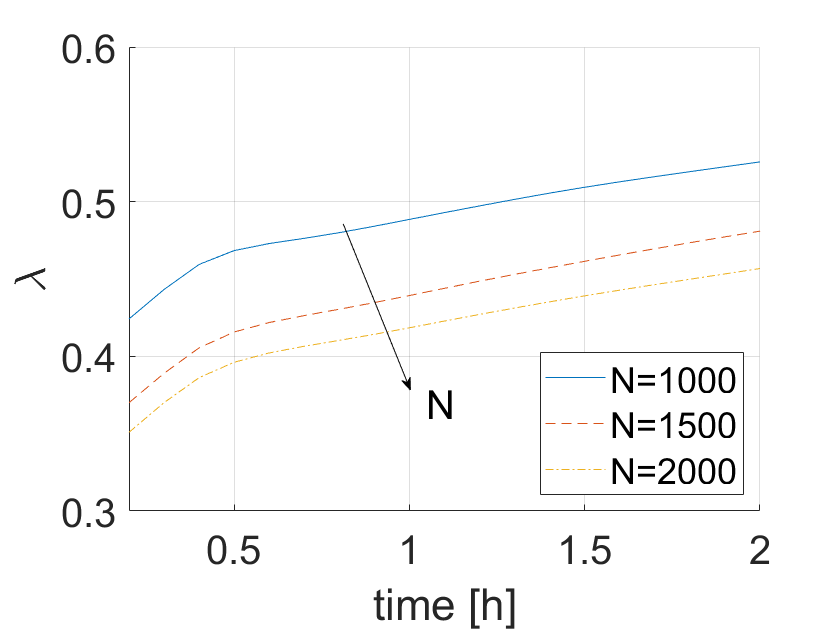}}
			\subfigure[\label{fig:iSR_poly}]
			{\includegraphics[width=0.49\textwidth]{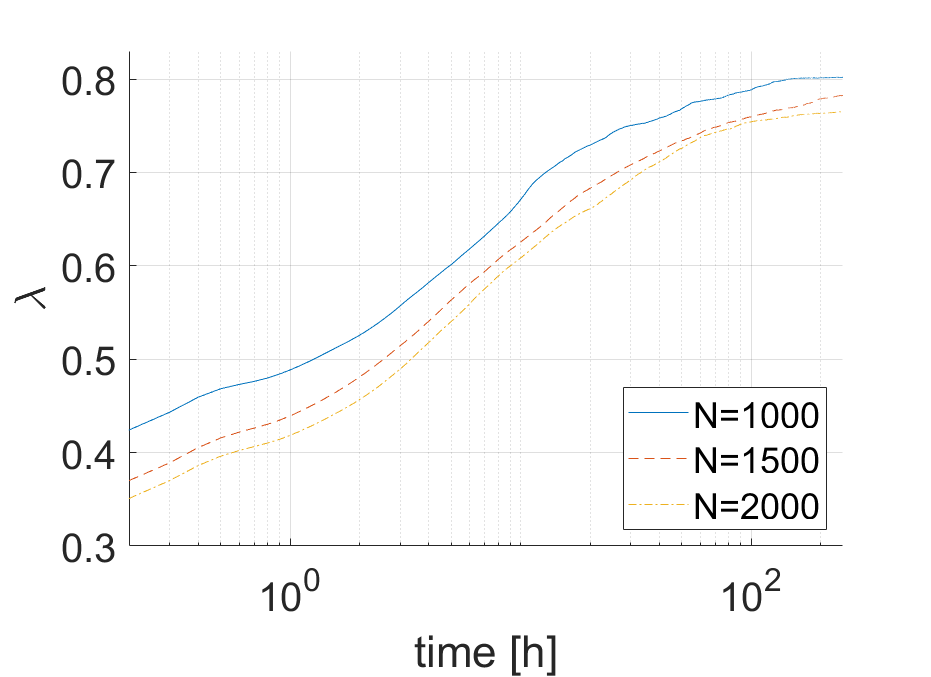}}
			\caption{Short and long-term evolution of inertial elongation $\lambda$ of the aggregate, as function of the number of bodies $N$ in the aggregate, for (a,b) monodisperse and (c,d) polydisperse particle size distribution. Long-term evolution plots (b,d) are shown using a semi-logarithmic (x) scale.}
			\label{fig:iSR_distrN}
		\end{figure*}
		
		First, we analyze the evolution of the aggregate as it reshapes towards an equilibrium condition under its own gravity. The overall shape of the aggregate can be monitored using its inertial elongation $\lambda$, as defined in Section~\ref{s_2:problem}. Figure~\ref{fig:FvsShape} shows the short- (2 hours) and long-term (250 hours, on a semi-logarithmic scale) time evolution of $\lambda$, for different levels of friction and using both angular bodies (Fig.~\ref{fig:FvsCH_z} and~\ref{fig:FvsCH}) and spheres (Fig.~\ref{fig:FvsSP_z} and~\ref{fig:FvsSP}). Results shown in Figure~\ref{fig:FvsShape} refer to aggregates with 1,000 monodisperse bodies only. Some considerations can be made based on long-term evolution graphs in Figure~\ref{fig:FvsCH} and~\ref{fig:FvsSP}:
		\begin{enumerate}[(i)]
			\item \label{en:4.1.1(i)} angular bodies contribute to reaching lower values of $\lambda$ compared to spheres at any friction level;
			\item \label{en:4.1.1(ii)} angular bodies contribute to reaching a steady equilibrium state earlier compared to spheres at any friction level;
			\item \label{en:4.1.1(iii)} both simulation sets show a substantially different dynamical behavior between cases with friction versus case with no friction;
			\item \label{en:4.1.1(iv)} both simulation sets show that, when friction is present, the value of $\eta$ does not play a relevant role to assessing the final shape of the aggregate. Counter-intuitively, in the long-term there is no monotonic trend between $\eta$ and $\lambda$ of the aggregate at equilibrium.
		\end{enumerate}
		Most importantly, results in Figure~\ref{fig:FvsShape} indicate clearly that substantial differences exist in the evolution of the aggregate due to the angular/spherical shape of particles. The reason for this is purely geometrical and mainly due to two effects: the number of points of contact between bodies and the interlocking mechanism. In particular, each pair of spheres has only one point of contact, while two angular bodies can have multiple points of contacts, including edges and surfaces. Dissipation forces related to friction are exchanged at points of contact and having more points means more forces acting on the bodies: friction is acting on a different scale. The second effect is due to the geometrical interlocking between angular bodies, which hinder their motion and does not occur between spherical bodies. These are indeed the reason behind (\ref{en:4.1.1(i)}) and (\ref{en:4.1.1(ii)}): an overall larger dissipation effect acts on angular bodies, which as a result, reach equilibrium before spheres and form more elongated aggregates.

		Considerations (\ref{en:4.1.1(iii)}) and (\ref{en:4.1.1(iv)}) are closely related to the role of friction in the simulation. To better investigate it, we focus on the initial phases of the reshaping process. Fig.~\ref{fig:FvsCH_z} and Fig.~\ref{fig:FvsSP_z} show the first few hours of simulation. These graphs show that, at the very beginning of the simulation, the trend between $\eta$ and $\lambda$ is monotonic both for spheres and for angular shapes, as intuition would suggest. However, on a longer time scale, the chaotic nature of contact interactions between bodies smooths these differences until the monotonic trend is lost. According to our results, and in agreement with results by \cite{Sanchez2016}, friction shall never be omitted when simulating a real-world scenario, since even a small dissipation produces a relevant difference in the dynamical outcome. On the other hand, the value of $\eta$ does not appear to be relevant on a long-time scale, especially for the case of angular bodies. Our results show that the relevance of $\eta$ depends on the time scale considered in the simulation. For a 2-hour simulation (short-term plots) there is a difference between having $\eta=0$, $\eta=0.2$ and $\eta\ge0.4$ (above this value we have very similar behaviors), while for a 250-hour simulation significant differences occur only between $\eta=0$ and $\eta\ne0$. In principle, given sufficient simulation time, this simplifies greatly the setup of the numerical problem, since it allows to avoid expensive parametric studies to select the value of $\eta$.

		We performed additional simulations to investigate the effect of a non-uniform size distribution and number of particles. We use a modified Zhang distribution (see Section~\ref{ss_3.3:parameters}) and we simulated scenarios with 1,000, 1,500 and 2,000 particles. As discussed, all simulations are performed under the same values of total mass $M$ and bulk density $\rho_b$ of the aggregate. For the case of polydisperse and higher number of particle simulations, material density and particle size are adjusted to match $M$ and $\rho_b$. All additional simulations are performed using angular bodies and a coefficient of friction $\eta=0.6$. Figure~\ref{fig:iSR_distrN} shows the time evolution of the inertial elongation $\lambda$ of the aggregate. Two effects are worth discussing.
		
		Our first considerations concern the settling time and time scales involved in the simulations. A meaningful metric we can use is the ``knee'' between the steep and mild slope of the curve. This is better visible in the short-term plots (Figures~\ref{fig:iSR_mono_z} and~\ref{fig:iSR_poly_z}) and identifies the transition point from an initial phase of fast settling to a phase of slower settling. In the case of monodisperse distribution (Fig.~\ref{fig:iSR_mono_z}) the knee occurs earlier for higher number of bodies, suggesting that a higher number of bodies contribute to shorten the fast settling phase. This is due to the increased surface interactions, which are the main source of motion dissipation. As expected, a higher number of contact interaction reduces the settling time. On the other hand, the same does not apply (or applies to a much lower extent) in the polydisperse case (Fig.~\ref{fig:iSR_poly_z}), where the knee occurs nearly at the same time for all cases. However, in these simulations the knee occurs earlier compared to monodisperse cases, for any N. This is consistent with the increased number of contact points in polydisperse simulations compared to monodisperse ones: smaller particles fill within the interstices of larger bodies, thus increasing the overall particle surface area at contact within the aggregate.
				
		The second aspect to be discussed concern the elongation $\lambda$ of the aggregate. In this case, the curves appear to be translated vertically. Also, they appear to converge to a single solution for higher N. At a careful analysis of the numbers, this appear to be motivated by the resolution of the aggregate. The term ``resolution'' here refers to the relative size between a single particle and the whole aggregate: the higher N, the better the resolution. In fact, when computing the shape of the aggregate, the effect of resolution is not negligible: for a 1000-body aggregate, the error due to resolution could be as high as 10-20\% (size of a single particle vs size of the aggregate). These errors are consistent with the vertical shift in the curves. Consistently, polydisperse fragments converges to a solution for higher N compared to monodisperse. In fact, the polydisperse mixture contains particles of bigger size (and thus have lower resolution) compared to the monodisperse ones (whose particle size is the mean of polydisperse distribution).
		
		As observed for the case of inter-particle friction, these effects are clearly visible, but limited to the first few hours of the numerical simulation. After that, the problem becomes highly chaotic and the effects of single parameters are not identifiable unequivocally any longer. In fact, at equilibrium, none of the aforementioned relations exist, except for the case of polydisperse simulations, where an increase in the number of bodies can be still correlated monotonically to a higher motion dissipation, and thus a more elongated final shape. Monodisperse simulations show a behavior similar to what already observed when comparing spheres with angular particles: aggregates that settle faster at the beginning of the simulation, reach equilibrium earlier compared to slower settling aggregates, which in the long term form more rounded shapes.

	\subsection{Final aggregate at equilibrium}

		\begin{figure}[t]
			\centering
			\includegraphics[width=0.49\textwidth]{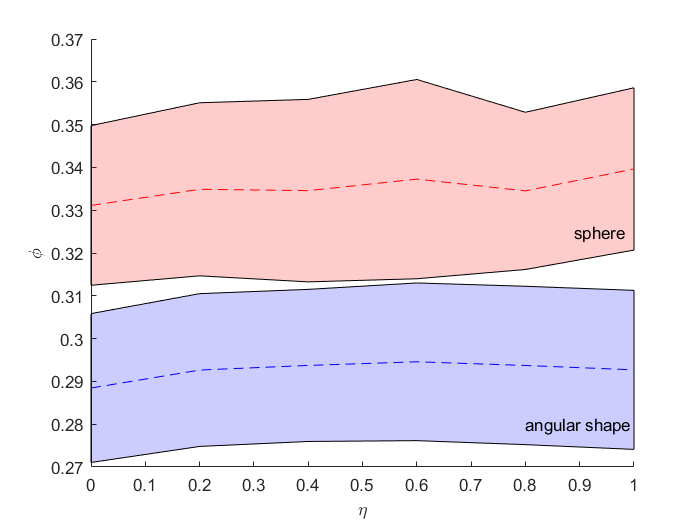}
			\caption{Porosity $\phi$ vs friction coefficient $\eta$ for final aggregates with spheres or angular bodies. The whole range is shown between minimum and maximum volume aggregate. Data refer to simulations with 1,000 monodisperse particles.}
			\label{fig:porosity}
		\end{figure}
		
		\begin{figure*}[t]
			\centering
			\subfigure[\label{fig:pack_d_CH}]
			{\includegraphics[width=0.49\textwidth]{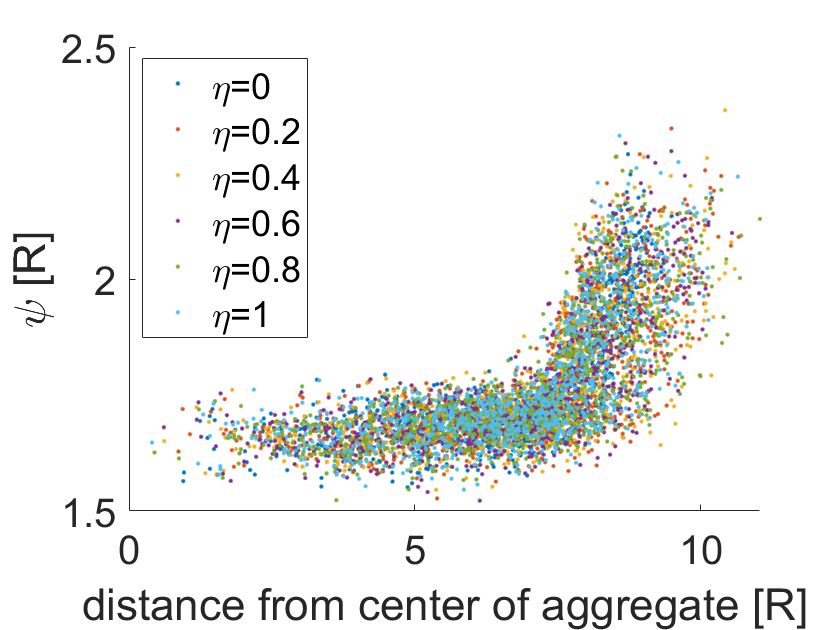}}
			\subfigure[\label{fig:pack_hist_CH}]
			{\includegraphics[width=0.49\textwidth]{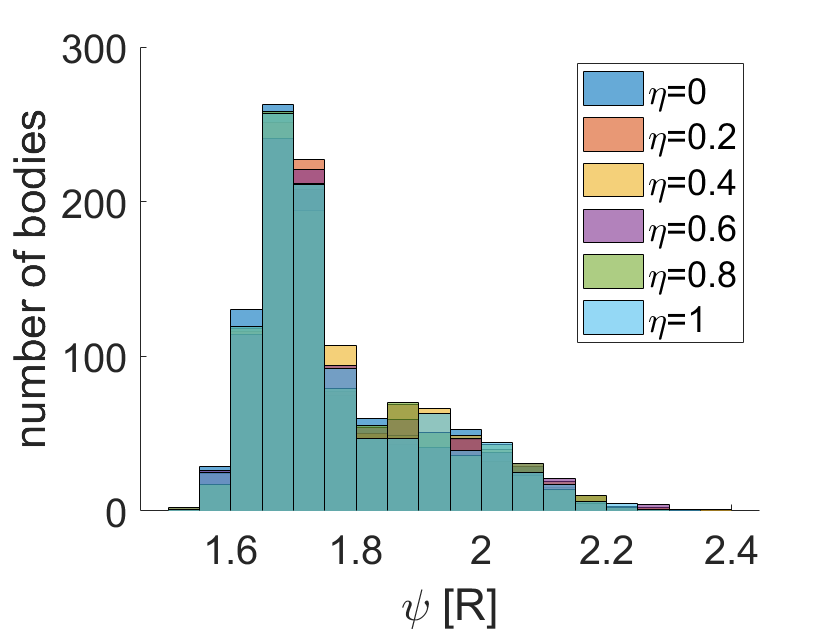}}
			\subfigure[\label{fig:pack_d_SP}]
			{\includegraphics[width=0.49\textwidth]{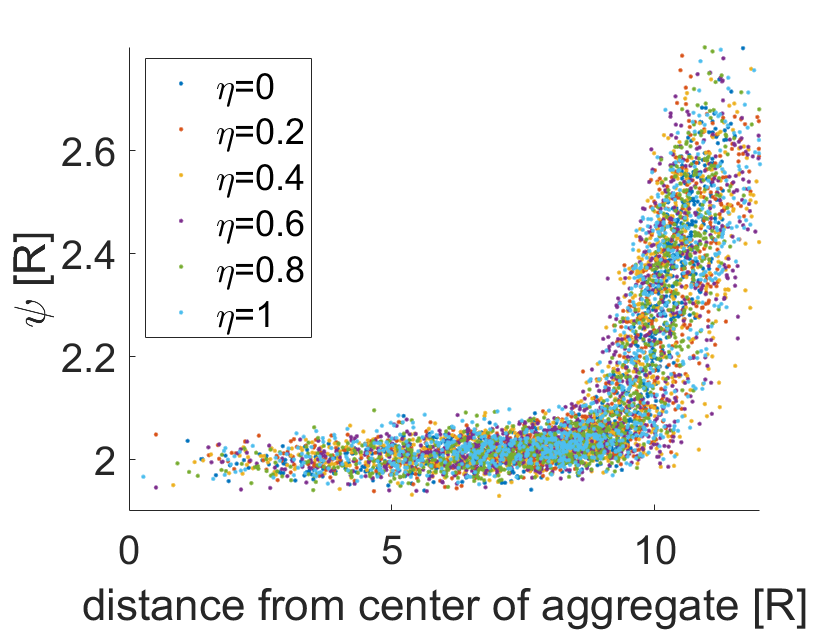}}
			\subfigure[\label{fig:pack_hist_SP}]
			{\includegraphics[width=0.49\textwidth]{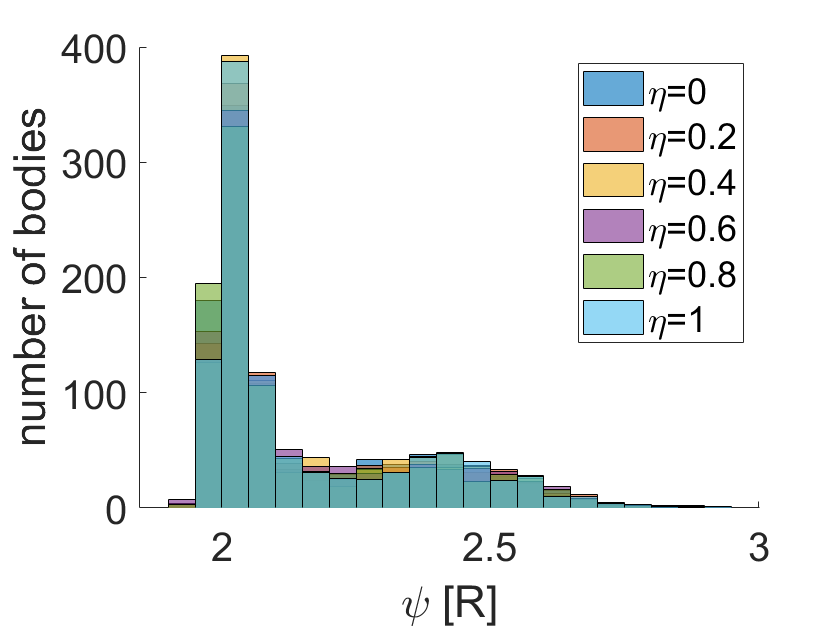}}
			\caption{Packing index $\psi$ for each body within the final aggregate, for different levels of friction coefficient $\eta$. Distribution as function of the distance of body from the center of the aggregate and histogram for (a,b) angular bodies and (c,d) spheres. Data refer to simulations with 1,000 monodisperse particles.}
			\label{fig:pack}
		\end{figure*}
		
		\begin{figure*}[t]
			\centering
			\subfigure[\label{fig:Fc_d_CH}]
			{\includegraphics[width=0.49\textwidth]{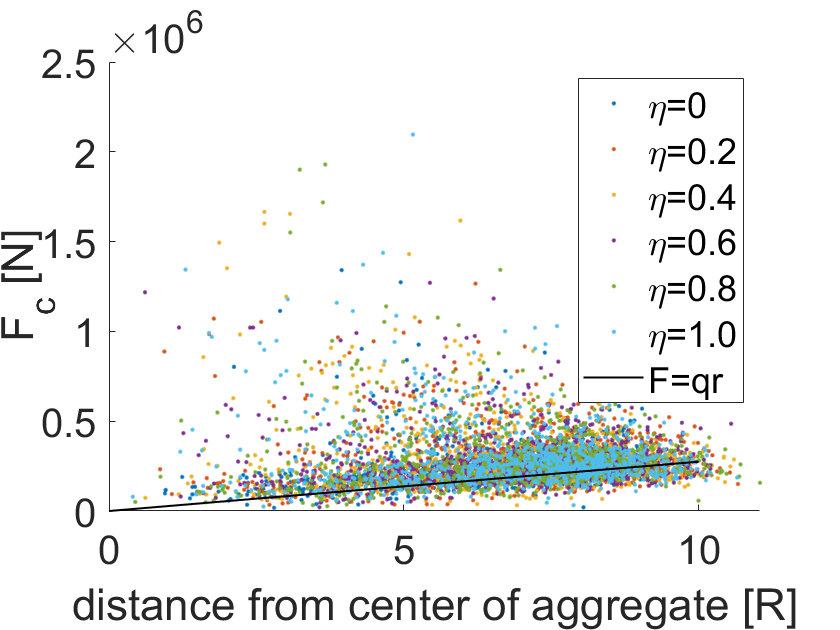}}
			\subfigure[\label{fig:Fc_hist_CH}]
			{\includegraphics[width=0.49\textwidth]{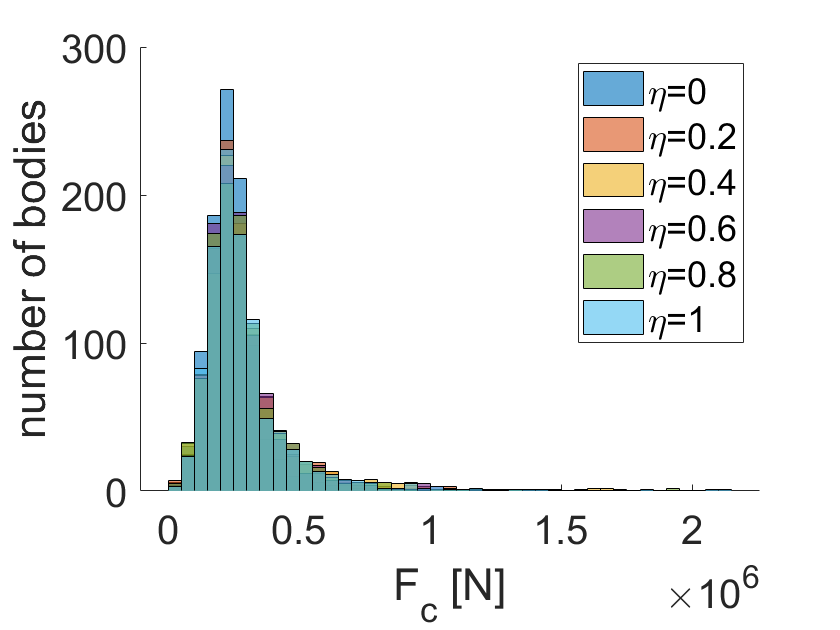}}
			\subfigure[\label{fig:Fc_d_SP}]
			{\includegraphics[width=0.49\textwidth]{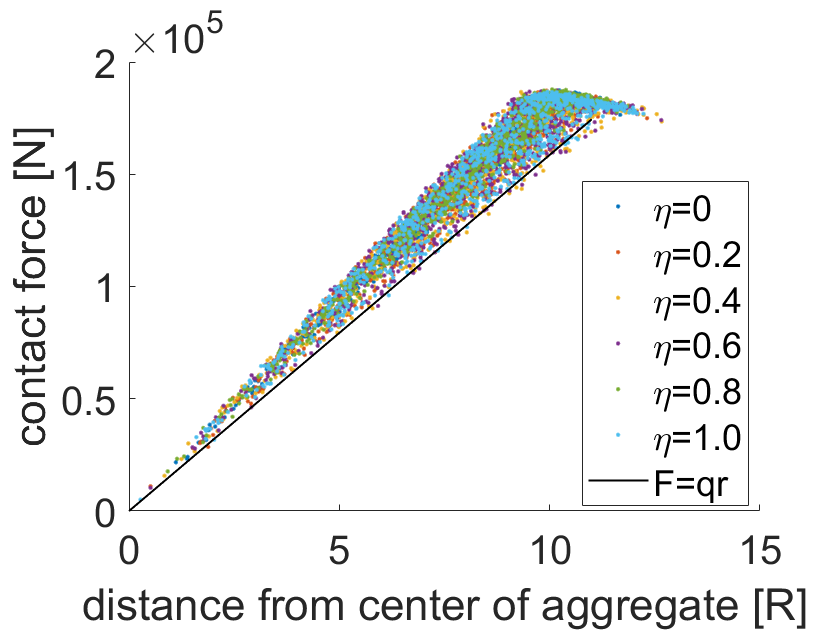}}
			\subfigure[\label{fig:Fc_hist_SP}]
			{\includegraphics[width=0.49\textwidth]{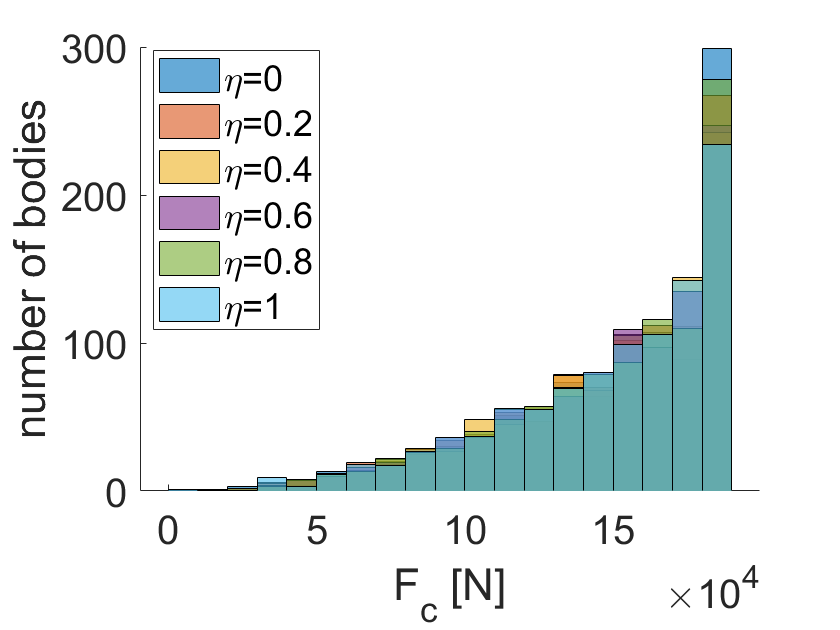}}
			\caption{Resulting contact force $F_c$ acting on each body within the final aggregate, for different levels of friction coefficient $\eta$. Distribution as function of distance of body from the center of the aggregate and histogram for (a,b) angular bodies and (c,d) spheres. Data refer to simulations with 1,000 monodisperse particles.}
			\label{fig:Fc}
		\end{figure*}
		
		After the transient phase, the aggregate reaches an equilibrium condition. We discuss here the properties and internal rubble-pile structure of the final aggregate. To provide quantitative means of comparison between simulations, we compute quantities defined in Section~\ref{s_2:problem} and discuss the results both from a global perspective, i.e.\ by considering the aggregate as a whole, and from a local perspective, i.e.\ by tracking each body in the aggregate. From a \textit{global} point of view, the results of simulations are used to constraint relevant quantities into ranges of values. This approach is motivated by the arbitrariness of the aggregate's surface definition, as discussed in Section~\ref{ss_3.5:finalaggr}. For each simulation and for each final aggregate, we compute the minimum and maximum volume aggregates. A global information of interest is the porosity $\phi$ or bulk density of the aggregate. Since the mass of the aggregate does not change between minimum and maximum volume aggregates, $\phi$ is minimum for minimum volume aggregates and maximum for maximum volume aggregates. Figure~\ref{fig:porosity} shows the estimate of porosity range for final aggregates made of spheres or angular bodies, for different levels of friction. In agreement with results on time evolution of the aggregate, friction plays a very minor role, while the shape of the particle affects significantly the result. As expected, the packing of angular bodies is more compact and efficient with respect to spheres and they occupy a higher fraction of total volume, thus leading to lower values of porosity ($\phi_{\text{ang}}\simeq0.29\pm0.02$ vs $\phi_{\text{sph}}\simeq0.33\pm0.02$).

		As \textit{local} metrics, we use the packing index $\psi$ associated to each particle and the resulting contact force $F_c$ acting on each particle. For both quantities, we look into their values and distribution within the aggregate. As defined in Section~\ref{s_2:problem}, $\psi$ is computed for each body in the aggregate as the mean distance (from center to center of the bodies) between the body and its 12 closest neighbors. All distances are normalized to $R=D/2$, where $D$ is the characteristic size of the angular body or the diameter of the sphere. Figure~\ref{fig:pack_d_CH} and~\ref{fig:pack_d_SP} show the packing distribution of the particles as function of their distance from the center of the aggregate for the case of, respectively, angular bodies and spheres. In both cases the distribution cloud has a plateau at lower distance from center and a steep ascent at higher distance. The plateau indicates a diffusely regular packing in the inner layers of the aggregates up to its external layers and surface, identified by the ascending part of the graph. Although sharing a similar behavior with spheres, angular bodies confirm to be more densely packed. Also, spheres appear to have a more regular packing in the inner layers, where all bodies are nearly at the condition of optimal packing ($\sim2$), while a more chaotic packing is observed on angular bodies due to their uneven and unequal shape. The same is confirmed in the histograms in Fig.~\ref{fig:pack_hist_CH} and~\ref{fig:pack_hist_SP}, where spheres show a larger discontinuity between the cluster around optimal packing conditions and the external layers. As in previous discussions, all graphs in Figure~\ref{fig:pack} confirm that the value of friction coefficient $\eta$ is not relevant to the local packing distribution of the aggregate.

		Similarly to the packing index, the contact force distribution among the particles can give useful insights on the internal structure of the aggregate. Contact forces are produced between bodies due to gravitational pulls acting on them. Ideally, for an aggregate at equilibrium, the net contact force on each internal particle equals the gravity force at the field point where the particle is. The aggregate is a distributed source of gravity and its internal gravity field can be approximated by the gravity field inside a solid shape. The gravity field inside a solid sphere follows the well-known relation
		\begin{equation}
			a_s(r)=\frac{GM_s}{R_s^3}r
		\end{equation}
		where $G$ is the universal gravitational constant, $M_s$ is the total mass of the solid sphere, $R_s$ is its radius and $r$ is the distance of between the field point and the center of the sphere. The gravitational force acting on a body of mass $m$ located inside the solid sphere at distance $r$ is linearly dependent on its distance from the center of the sphere
		\begin{equation}
		\label{eq:F_ssphere}
			F_s(r)=qr	\qquad \text{with}	\qquad q=\frac{GM_s m}{R_s^3}
		\end{equation}	
		In our case, the final aggregates are not perfectly spherical but their elongation is small and Eq.~\eqref{eq:F_ssphere} can be used as a comparison mean against force distribution inside the aggregate. In particular, the linear coefficient $q$ is computed for two sample spheres, representative of aggregates with angular and spherical particles. These consider the total mass of the aggregate $M=M_s$, the mass of each body $m$ (we consider here cases with 1,000 monodisperse particles, where both $M$ and $m$ are equal for all simulations) and the mean radius of the aggregate $R_{\text{agg}}=R_s$. The mean radius of the aggregate $R_{\text{agg}}$ is computed from its mean volume (between maximum and minimum volume aggregate). Since aggregates with angular particles are smaller in size, $q_{\text{ang}}>q_{\text{sph}}$ and forces inside them are expected to be higher. This is confirmed by results of simulations in Figure~\ref{fig:Fc}. Figure~\ref{fig:Fc_d_CH} and~\ref{fig:Fc_d_SP} show the distribution of forces inside the aggregate, as function of the distance between the body and the center of the aggregate. As expected, the distributions behave linearly with $r$ and are coherent with the gravity field inside the sample solid sphere (black line). Based on Fig.~\ref{fig:Fc}, the choice of particle shape appears to be very relevant to assessing the distribution of contact forces within the aggregate. In the case of spheres, the distribution of forces is extremely regular, suggesting regular internal packing, while angular bodies show a more chaotic distribution. The chaotic nature of interactions between angular bodies, due to a higher number of contact points and to the interlocking mechanism, causes very high forces on a number of particles in the inner layers of the aggregate. This does not happen in the regularly packed spheres, where highest forces are found on the external surface of the aggregate. High forces appear to be caused by the angularity of contacts involving vertices and edges and are not observed on the smoother spherical surfaces. A more detailed analysis would be beneficial to fully investigate this behavior: we highlight it here as an interesting point for a follow-up analysis, which is however out of the scope of the current work. The regularity/irregularity of force distributions are in agreement with results on packing index shown in Figure~\ref{fig:pack}. Again, the value of $\eta$ does not play a relevant role.

		No relevant differences are observed in the internal packing distribution when dealing with a higher number of particles. Instead, as expected, aggregates with polydisperse material have smaller porosity (5-15\% smaller with respect to monodisperse aggregates). In this case, smaller particles fill the voids between bigger particles and help avoiding the formation of crystallized packing.

	\subsection{Considerations on the angle of friction}
	\label{ss_4.3:continuum}
		When dealing with self-gravitating objects, the classical approach is to compare their shape to hydrostatic sequences of equilibrium. The theory of continuum shows how self-gravitating fluids follow minimum energy configurations (as Jacobi and MacLaurin sequences), which correlate their angular momentum or spin with their shape~\citep{Chandrasekhar1969}. Our simulations show equilibrium configurations of non-rotating aggregates. In the context of the continuum model, the equilibrium figure of a non-rotating fluid is a perfect sphere. However, in our simulations, the kinetic energy of the system is such that the granular material does not behave as a fluid, but rather as a granular solid. In fact, shapes of rubble piles can be very different from hydrostatic equilibrium ones~\citep{Minton2008,Harris2009,Scheeres2015} and no direct comparison is possible. In general, self-gravitating granular aggregates relax towards a more spherical shape but, even those formed by frictionless particles, never reach precisely the hydrostatic spherical shape~\citep{Tanga2009a,Sanchez2012}. This is due to mechanisms that hinder the motion of particles within the granular solid. To provide a better theoretical model, capable of dealing with this phenomenon, Holsapple has applied the continuum theory to non-fluid (solid) bodies. In this effort, he extended the range of configurations attainable by a self-gravitating object by introducing the effects of a non-zero angle of friction \cite[e.g.][]{Holsapple2001,Holsapple2004,Holsapple2007,Holsapple2010}. 
		
		The angle of friction is a quantity not directly related to surface friction, but it is rather a measure of the motion-hindering effects occurring within the interior of the object. In a direct comparison with a granular system, it would include all such mechanisms related to contact interactions. As shown by~\citep{Holsapple2010}, the range of admissible shapes of a self-gravitating ellipsoid is greatly enhanced by a non-zero angle of friction. For example, for a friction angle of 30 deg (which is rather common in soil-like material), a non-rotating ellipsoid can attain almost any shape, except for extremely elongated ones, when c/a is very close to zero~\citep[see e.g.\ Fig.1d in ][]{Holsapple2010}. However, although this model is certainly more accurate than using simpler fluid equilibrium sequences, it still relies on major simplifications of the granular problem. In particular, the model implies that the equilibrium shape attained by an object depends only on its angular momentum/spin and friction angle. In principle, this is true for granular systems as well. However, the friction angle of granular media is strongly dependent on the kinetic energy of the particles: according to its energy level, the granular media can behave as a gas, fluid or solid, each having a very different friction angle. To demonstrate it, we provide a simple example, considering two different dynamical scenarios.
		
		The first scenario considers an initially dispersed cloud of particles, collapsing under self-gravity. In this case, spherical (or nearly spherical) shapes can be attained for the aggregate at equilibrium~\cite[see e.g.][]{Tanga2009}. The greater contribution to the aggregate's final shape is provided by the initial phase, where particles follow free gravitational dynamics, while contact interactions contribute marginally to it. In this first phase, the dispersed particles act as a granular gas and then transition to a granular fluid as soon as the contacts becomes more numerous. In these initial phases, the angle of friction is extremely low, and this allows to minimize the motion-hindering mechanisms that act against reaching hydrostatic equilibrium. This is what we obtain in our simulations of parent aggregate formation (Figure~\ref{fig:parent_sim}), as shown in Section~\ref{ss_3.4:initialaggr}. Consistently, we form nearly spherical aggregates (with $\lambda\simeq 1$), as shown in Figures~\ref{fig:bigagg} and~\ref{fig:bigagg_poly}. 
		
		The second scenario considers the same particles, but arranged in a different initial configuration, such that simulations starts from an already formed aggregate. After settling, the aggregate reaches the form of a granular solid, with a much higher angle of friction compared to the gaseous and fluid phases. This makes a substantial difference, since particles are experiencing contact interactions and interlocking from the beginning. They are not free to move in the entire space but only to roll/slide over each other. In this case, the initial shape of the aggregate strongly biases the final shape achieved after the reshaping process, which is no longer close to spherical. So, unlike the fluid case and the Holsapple case, the granular problem does not admit a unique equilibrium solution. The shape at equilibrium still depends on its angular momentum and angle of friction of the granular media, but also on the complex contact history of the aggregate. 
		
		In our case study, this refers mainly to the way the initial aggregate is created and how the numerical simulation is initiated. Since we are not studying the failure conditions of our aggregates, the approach of Holsapple cannot be used directly to infer the theoretical angle of friction of our ellipsoidal aggregates based on their axes ratio. However, this approach can provide useful information to constraint the angle of friction of our aggregate. Holsapple investigates the limiting failure conditions of objects, using either the Drucker-Prager or the Mohr-Coulomb failure criterion. These are based on the idea that failure is due to an exceeding shear stress. This limiting condition provides a lower limiting value for the angle of friction: regardless of the contact history of the aggregate, we can state that a certain shape (value of semiaxes of the ellipsoid) implies that the angle of friction can never be less than the value found with the continuum model by Holsapple. Should it be lower than that, the aggregate would have been further relaxed towards hydrostatic equilibrium. In particular, we refer here to the Drucker-Prager (DP) criterion, applied to ellipsoidal shapes, and we look for its limiting condition (we do not consider any cohesive term):
		\begin{equation}
		\label{eq:DP}
			\sqrt{J_2}+3sp=0
		\end{equation}
		where $J_2$ is the second invariant of the deviator stress tensor, $p$ is the hydrostatic pressure (or mean normal stress) and $s$ is a constant that can be related to the angle of friction $\theta_f$ of the aggregate. Respectively, they can be expressed as:
		\begin{align}
			&J_2=\frac{1}{6} \left[ (\sigma_1-\sigma_2)^2 + (\sigma_1-\sigma_3)^2 + (\sigma_2-\sigma_3)^2 \right] \\
			&p=\frac{1}{3}(\sigma_1+\sigma_2+\sigma_3) \\
			\label{eq:s_phi} &s=\frac{2 \sin(\theta_f)}{\sqrt{3}(3-\sin(\theta_f))}
		\end{align}
		where $(\sigma_1,\sigma_2,\sigma_3)$ are the principal stresses. We remark that the expression for $s$ is not unique~\citep{Chen1988}: we use Eq.~\eqref{eq:s_phi} to be consistent with results obtained in~\citep{Holsapple2004,Holsapple2010}. To have a meaningful means of comparison with the continuum model, we apply the DP criterion to an equivalent ellipsoid, built upon our final rubble-pile aggregate. The equivalent ellipsoid is defined as the one having the same principal moments of inertia of the rubble-pile aggregate. In the literature this is often referred to as the Dynamically Equivalent Equal Volume Ellipsoid (DEEVE). The use of an equivalent ellipsoid for our case study is justified by the fact that we use ellipsoidal aggregates: all of their equilibrium shapes can be well approximated using ellipsoids. To find the semi-axes of the equivalent ellipsoid, we first compute the inertia tensor and principal moments of inertia of the rubble pile aggregate. When performing this operation, we consider the full inertia tensor of each single particle. The semi-axes of the equivalent ellipsoid are then computed from principal inertia moments, as if they belonged to a homogeneous ellipsoid. The principal stresses of the ellipsoid can be expressed in terms of its gravity potential terms. The gravity potential $U$ of an ellipsoid can be written in a co-rotating frame (rigidly rotating with the body) as~\citep{Chandrasekhar1969}:
		\begin{equation}
			U=\pi\rho G (- A_0 + A_x x^2 + A_y y^2 + A_z z^2)
		\end{equation}
		where $\rho$ is the homogeneous density of the ellipsoid, $G$ is the universal gravitational constant and terms $(A_0,A_x,A_y,A_z)$ depend only on the semi-axes of the ellipsoid. We evaluate the stress tensor at the center of a non-rotating ellipsoid, which can be written as~\citep{Holsapple2010}:
		\begin{equation}
			\mathbf{\sigma}_c=-\pi\rho^2 G
			\begin{bmatrix}
			A_x & 0 & 0 \\ 0 & A_y & 0 \\ 0 & 0 & A_z
			\end{bmatrix}
			\begin{bmatrix}
			a^2 & 0 & 0 \\ 0 & b^2 & 0 \\ 0 & 0 & c^2
			\end{bmatrix}
		\end{equation}
		All terms in Eq.~\eqref{eq:DP} can be expressed in terms of the semi-axes of the ellipsoid, which we denote with $a>b>c$. Finally, we can find the angle of friction $\theta_f$ that satisfies Eq.~\eqref{eq:DP}, as a function of semi-axes only.

		\begin{figure}[t]
			\centering
			\includegraphics[width=0.5\textwidth]{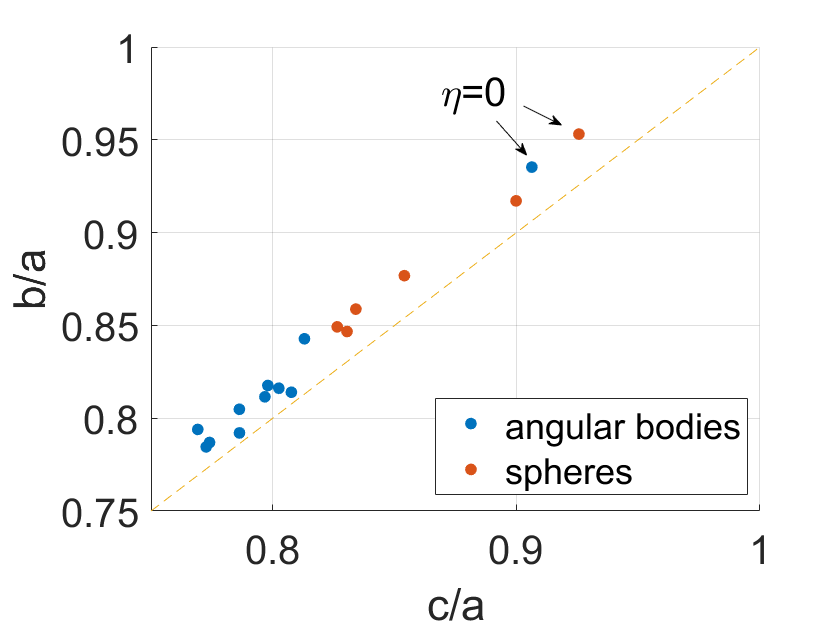}
			\caption{Shape of equilibrium aggregates in terms of semi-axes of equivalent ellipsoid, with $a>b>c$. Particle shape and cases with no surface friction $\eta=0$ are highlighted.}
			\label{fig:semiaxes}
		\end{figure}
		
		Figure~\ref{fig:semiaxes} shows the shape of final aggregates at equilibrium. All of them result close to the b/a=c/a line, i.e.\ nearly prolate ellipsoids. This is clearly biased by the choice of the initial shape, which is a prolate ellipsoid (with b/a=c/a=0.4) for all simulations. As discussed this prevents us from drawing general conclusion based on their final shape only, but it allows to derive a lower limiting value for their angle of friction. If excluding cases with no friction (marked with $\eta=0$ in the figure), the semi-axes are in the range [0.83-0.92] for aggregates with spherical particles and [0.77-0.84] for angular bodies. These correspond to minimum angles of friction of about 4 deg for the spheres. As expected, this value is consistent with the findings of \cite{Tanga2009}, when one considers in their article the evolution of the ellipsoid having similar flattening and the lowest angular momentum. 
		The interesting difference that we find is the friction angle for angular bodies, around 6 deg, which denote a modest but relevant (+50\%) increase provided by the angularity of particles. 
		
		An accurate evaluation of the angle of friction of the aggregates is out of the scope of this paper, which is rather focused on highlighting the role of surface parameters and particle shape in the dynamical reshaping process. Also, we recall here the results of preliminary tests discussed in Section~\ref{ss_3.2:tests}, which are performed using the same surface parameters used in ellipsoid simulations. In particular, the angle of slope test provides an upper limit in terms of slope attainable by the granular medium, which are observed up to values of about 65 deg.

		In this context, we highlight that a more comprehensive treatment of the problem would consider many other degrees of freedom when modeling the rubble-pile aggregate. A very relevant one concerns the internal structure of these bodies. Elongated shapes might be sustained by inhomogeneous internal structures (or density distributions). To provide few examples, \cite{Lowry2014} speculated about the interior of the highly elongated asteroid Itokawa based on its global shape and spinning dynamics: according to them a strong inhomogeneity within Itokawa (two separate bodies with different density) could possibly explain their measures of spin dynamics. \cite{Kanamaru2019} also proposed an inhomogeneous model for Itokawa, to cope with its gravitational potential and surface properties. In this respect, our approach is clearly simplified and do not address the problem in its whole complexity (we make the bulk hypothesis that such bodies have a full rubble-pile structure with homogeneous density). However, our ultimate goal is not to reproduce the high fidelity internal structures, but rather to exploit the rubble-pile structure to reproduce the self-gravitational dynamics of these objects.

\section{Conclusions}
\label{s_5:conclusion}
	In this paper we study the role of particle shape and surface/contact parameters in N-body simulations, considering the full gravity-contact problem. We discuss the results of test scenarios of granular dynamics to quantify and qualitatively characterize the angle of slide and angle of repose of a realistic granular media, made of angular particles. As a case study, we reproduce the natural reshaping process of an elongated ellipsoidal rubble-pile object by numerical simulations. We investigate the effects of simulation parameters, including particle shape, surface friction, particle size distribution and number of particles, and discuss their contribution to the overall angle of friction within the granular medium. We summarize here the main outcomes of the paper:
	\begin{itemize}
		\item \textit{Friction.} The presence/absence of friction is very relevant, but not its value. This is supported by results in all simulations and test scenarios. Friction is observed at a different scale for the case of angular bodies compared to spheres. This is due to the increased number of contact points (many, compared to only two per sphere couple) and enhanced by the chaotic nature of interactions between angular bodies. The overall effect is that angular bodies contribute to reaching equilibrium earlier than spheres and, in the long time, the value of friction does not affect the final shape of the aggregate.
		\item \textit{Porosity.} As expected, aggregates with spheres have higher porosity, while packing of angular bodies is more efficient in terms of volume occupied. This appears to be a purely geometrical feature of the granular medium and has direct consequences on the dynamics and evolution of rubble pile objects. However, based on this consideration only, it is hard to speculate about global properties of small celestial bodies and compare with their high porosity values, due to the lack of data about their internal structure. This is the missing ingredient we would need to formulate hypotheses of correlation between local (internal) geometrical structure and global properties.
		\item \textit{Final shape of aggregate.} Aggregates with spheres are more rounded and closer to hydrostatic equilibrium shape. Also, their reach equilibrium later in time. This is because they move freely on the surface, due to a lower friction (less contact points) and to the absence of mechanisms to hinder their motion such as geometrical interlocking, which instead play a relevant role for the case of angular bodies.
		\item \textit{Aggregate internal structure.} Spheres have a more regular packing as they form ordered crystal structures. Angular bodies have a more chaotic packing structure. This has consequences on the internal properties of the aggregate: a clear example is shown on its internal force distribution, which is very different in the two cases.
	\end{itemize}
	
	While assessing the role of particles in N-body simulations, we remark that these results are valid for the specific scenario studied, where all particles have equal density within a homogeneous aggregate that slowly settles under its own gravity, and with no other external actions. As mentioned for the case of porosity, any direct generalization of these results to the case of small celestial bodies would be possible provided that we have a better understanding and data available on the internal structure of such bodies. This work is part of a basic research effort, aimed at investigating the dynamical behavior of gravitational aggregates. In this context, general results and considerations clearly emerge.
	
	Angular bodies are certainly more realistic than spheres since they are capable to simulate to a more complex extent the contact interaction mechanism. The results give a clear indication towards the use of angular bodies for a better understanding of the dynamics involving the internal structure and global properties of rubble piles asteroids.
	
	A further argument of support is provided by the analysis of short- versus long- term dynamics. We observe that, although clear trends may exists within the first few hours of simulation, most of them disappear in the long term. For instance, considering reshaping process, we observe that after two hours of simulation clear trends exist between the shape of the aggregate and simulation parameters: in this case a more elongated aggregate is produced by (i) higher surface friction between particles, (ii) higher number of fragments (higher resolution of the aggregate) or (iii) non-uniform size distribution of particles. As expected, these three phenomena contribute to hindering the motion of particles within the aggregate, either by a direct increase of surface friction (i), or by its indirect increase after enlarging the surface area at contact (ii,iii). However, in all cases, when looking at results at equilibrium (after 250 hours of simulations), these trends are no more observed. This is interpreted as an effect of the extremely chaotic dynamical environment, which on the long-term dominates over the effects of simulation parameters. The only effect that survives in the long term is that of particle shape, which is observed in all simulations. 
	
	Our simulations indicate clearly that relevant differences exist when using angular bodies compared to using spheres. Despite the importance of spherical particles in disclosing the physics of granular materials, both in laboratory and numerical simulations, our results highlight that relevant differences exist when irregular fragments are adopted. The additional level of complexity that non-spherical shapes bring in numerical simulations, appears to be an unavoidable step to better reproduce real-world scenarios.

\section*{Acknowledgment}
	This project has received funding from the European Union's Horizon 2020 research and innovation programme under the Marie Skłodowska-Curie grant agreement No 800060. Part of the research was carried out at Laboratoire Lagrange, Observatoire de la C\^ote d'Azur, acting as secondment institution of the Marie Skłodowska-Curie project GRAINS.
	Part of the research work was carried out at the Jet Propulsion Laboratory, California Institute of Technology, under contract with the National Aeronautics and Space Administration.
	F.F. would like to thank Anton Ermakov for the fruitful discussions that inspired the setup of the angle of slide test scenario.
	The authors would like to thank the anonymous reviewers for their comments and suggestions that helped to increase the quality of the paper.

\bibliographystyle{elsarticle-harv}
\bibliography{references}

\end{document}